\documentclass[11pt,a4paper]{article}

\usepackage{jheppub}
\usepackage{amsmath,amssymb,mathtools,bm,amsthm}
\usepackage{mathrsfs}
\usepackage{microtype}

\theoremstyle{plain}

\theoremstyle{remark}

\preprint{UT-WI-33-2026}
\title{Dirichlet gravitons in AdS$_4$ and
$\mathcal{L}_{\Lambda}w_{1+\infty}$ wedge algebra}
\author{Hare Krishna}

\affiliation{Weinberg Institute, Department of Physics, University of Texas at Austin, Austin, TX 78712, USA}

\emailAdd{hkrishna.phy@gmail.com}

\abstract{
We derive the AdS deformation of the flat-space soft-graviton algebra with Dirichlet boundary conditions. Starting from the AdS$_4$ spinor-helicity representation, we Mellin-transform linearized graviton solutions and construct the 
AdS wedge, including its Laurent completion. The Dirichlet boundary condition pairs opposite helicities,
while AdS covariance fixes their relative normalization. Assuming linear closure and normalizing the global modes to reproduce the geometric AdS isometries, covariance and the Jacobi identities uniquely determine
the soft-mode bracket. The result is the wedge of
$\mathcal L_\Lambda w_{1+\infty}$, where $\Lambda=-\ell^{-2}$.
This provides a bulk derivation of the cosmological-constant deformation that incorporates the AdS boundary condition, without relying on collinear splitting functions or a self-dual truncation.
}

\begin{document}
\maketitle

\section{Introduction}
\label{sec:introduction}

Soft gravitons illuminate an infinite-dimensional symmetry structure underlying four-dimensional gravitational scattering. The universality of the leading and tree-level subleading soft graviton theorems \cite{Weinberg:1965,CachazoStrominger:2014} finds a natural explanation in the asymptotic symmetries of the gravitational $\mathcal{S}$-matrix. The leading theorem is equivalent to the Ward identity for BMS supertranslations \cite{HeEtAl:2015}, while the subleading theorem follows from a superrotation Ward identity \cite{KapecEtAl:2014}. In four dimensions, this connection extends to the universal logarithmic soft terms \cite{SahooSen:2019,KrishnaSahoo:2023,Krishna:2024}, which can also be derived from superrotation Ward identities \cite{AgrawalEtAl:2024}. Universality becomes more restricted at higher orders: already at sub-subleading order, the soft theorem contains a universal contribution together with theory-dependent corrections \cite{LaddhaSen:2017}. The celestial conformal basis provides a useful framework for organizing the universal structure.\\

A Mellin transform with respect to the external energies replaces
momentum eigenstates with states of definite conformal dimension
\cite{PasterskiShao:2017,PasterskiShaoStrominger:2017},
translating the successive powers of soft expansion into poles at
$\Delta=1,0,-1,\ldots$. Conformally soft limits promote soft gauge
and gravitational modes to celestial currents
\cite{DonnayPuhmStrominger:2019,AdamoMasonSharma:2019}.
The corresponding conformally soft graviton operators generate an
infinite tower of chiral currents which organize in the wedge algebra
of $w_{1+\infty}$
\cite{GuevaraEtAl:2021,Strominger:2021,
FreidelPranzettiRaclariu:2022,HimwichPate:2024}.
The algebra also survives the one-loop-exact quantum corrections of
self-dual gravity \cite{BallEtAl:2022}. Related phase-space and scattering constructions realize the same symmetry through higher-spin gravitational charges and their action on gravitational scattering states \cite{FreidelPranzettiRaclariu:2022,HimwichPate:2024}. Requiring consistency of the soft current algebra and associativity of the celestial operator product expansion constrains bulk interactions and scattering amplitudes \cite{MagoEtAl:2023,RenEtAl:2022}. The problem of consistently combining the two graviton helicity sectors has also been studied from the covariant phase-space perspective \cite{PranzettiSalluce:2026}. This also sheds some light on the putative celestial dual of gravity in asymptotically flat space. These celestial duals are more exotic, and we don't understand them well. \\

These developments also motivate a closely related question in the more established setting of AdS/CFT. What is the soft graviton algebra in the presence of a cosmological constant, and how is it realized in the boundary theory? Soft limits in curved backgrounds have been studied directly in (A)dS amplitudes and correlators using Mellin-momentum bootstrap and conformal Ward identities \cite{MeiMoBootstrap:2024,ChowdhuryEtAl:2024,MeiMo:2025}. Related de Sitter analyses have connected soft graviton limits to near-horizon symmetries and cosmological-constant corrections to supertranslation Ward identities \cite{MaoZhou:2024,ChattopadhyaySolanki:2026}. Taylor and Zhu proposed a cosmological-constant deformation, $\mathcal{L}_{\Lambda}w_{1+\infty}$, which incorporates the AdS$_4$ isometry algebra for negative $\Lambda$ \cite{TaylorZhu:2024}. Recently, Strominger and Wei constructed a realization of this algebra in CFT$_3$, generated by the averaged null energy operator, its conformal descendants, and their commutators \cite{StromingerWei:2026}. Further CFT$_3$ realizations and the associated celestial representation theory have since been investigated in ref.~\cite{Zhu:2026}. This identifies a concrete boundary realization of the deformed soft symmetry through the algebra of light-ray operators. Complementary bulk constructions give further insight into this deformation. Lipstein and Nagy identified a deformed kinematic algebra in self-dual AdS$_4$ gravity and extended it to a deformation of $w_{1+\infty}$ \cite{LipsteinNagy:2023}. Bittleston et al.\ recovered the $\mathcal{L}_{\Lambda}w_{1+\infty}$ algebra from the cosmological-constant deformation of the Poisson structure on twistor space \cite{BittlestonEtAl:2024}. More recently, Goodenbour et al.\ constructed the corresponding Hamiltonian charges from boundary ambitwistor data and related their spacetime representatives to CFT$_3$ light-ray operators \cite{GoodenbourEtAl:2026}. Higher-spin extensions of the deformed holographic symmetry algebra have also been considered \cite{BanerjeeEtAl:2026}. On the gravitational phase-space side, Di Giacomo et al.\ constructed higher-spin charges realizing $\mathcal{L}_{\Lambda}w_{1+\infty}$ on a restricted asymptotically (A)dS$_4$ phase space with boundary conditions allowing gravitational flux \cite{DiGiacomo:2026}. Together with the boundary light-ray construction and its gauge-theory precursor \cite{ShetaEtAl:2026}, these results connect the algebra to several complementary descriptions of bulk and boundary dynamics.\\

A bulk realization must account for the boundary conditions. Standard Dirichlet conditions reflect positive-helicity gravitons into negative-helicity gravitons, precluding a purely self-dual radiative sector \cite{StromingerWei:2026}. Di Giacomo et al.\ recently constructed higher-spin gravitational charges that realize the deformed algebra on a restricted phase space with boundary conditions allowing gravitational flux, working to first order in $\Lambda$ and quadratic order in the fields \cite{DiGiacomo:2026}. This is \textit{not} the standard Dirichlet boundary condition in AdS.\\

In this article, we construct the classical soft-mode algebra of AdS$_4$ gravitons with Dirichlet boundary conditions. Our starting point is the AdS spinor-helicity formalism of Nagaraj et.al.~\cite{NagarajPonomarev:2019,NagarajPonomarev:2020}. A Mellin transform organizes the linearized solutions into families of definite homogeneity, from which we construct a mode space using the AdS isometries. We retain both graviton helicities. This is essential because reflection at the AdS boundary exchanges the two helicities. Fixing the boundary metric relates their boundary data, while compatibility with the AdS isometries determines how the corresponding modes pair across the soft tower. More explicitly, Dirichlet boundary conditions pair the two graviton helicities by fixing the boundary metric and hence setting the linearized magnetic Weyl data to zero. Since the positive and negative helicity sectors satisfy $\mathcal B^{(\pm)}=\pm i\mathcal E^{(\pm)}$, their magnetic contributions cancel when their electric boundary data agree. A Dirichlet graviton combines opposite-helicity modes with matched electric data. In our mode expansion basis, the partner of $w^p_{\bar m,m}$ is $\bar w^{,3-p}_{\bar m,m}$. Requiring this pairing to respect the AdS isometries relates the relative coefficients at neighboring levels through the action of the transvections, fixing their dependence on $p$. Choosing equal electric normalization at $p=\frac32$ gives the Dirichlet-compatible combination $W^p_{\bar m,m}=w^p_{\bar m,m}+\Lambda^{3-2p}\bar w^{,3-p}_{\bar m,m}$. Thus, the boundary condition specifies which helicity data must match, while AdS covariance propagates their relative normalization throughout the tower.\\

We then determine a bracket on the Dirichlet-compatible modes by requiring antisymmetry, linear closure, and the Jacobi identity, together with the known geometric action of the AdS generators. Within the specified mode space and these assumptions, the bracket is uniquely fixed. The resulting algebra is the $\mathcal L_\Lambda w_{1+\infty}$ wedge. This symmetry algebra has already appeared in the boundary construction of Strominger and Wei ~\cite{StromingerWei:2026}. Our contribution is a bulk derivation of this known algebra that incorporates Dirichlet boundary conditions and makes the pairing of the two helicities explicit. Our construction is the classical Lie algebra of linear soft modes. Its extension to nonlinear gravitational charges remains an open question.

\paragraph{Notation and conventions.}
We distinguish the Einstein cosmological constant from the
parameter entering the algebra
\begin{equation}
    \Lambda_{\rm Ein}=-\frac{3}{\ell^2},
    \qquad
    \Lambda\equiv\frac{\Lambda_{\rm Ein}}{3}
    =-\frac{1}{\ell^2}.
    \label{eq:Lambda-def}
\end{equation}
The representation-theoretic calculations use the complexified
AdS isometry algebra, with $z$ and $\bar z$ treated as independent
projective coordinates. A bar on $\bar w$ labels the opposite
helicity sector.


\section{Mellin-homogeneous AdS$_4$ graviton modes}
\label{sec:ads_mellin_gravitons}
We construct a family of linearized AdS$_4$ graviton modes from the
spinor-helicity solutions of Nagaraj et al.~
\cite{NagarajPonomarev:2019,NagarajPonomarev:2020}.  For the
algebraic construction, we complexify the spinor variables and treat
the dotted and undotted spinors as independent.  This makes manifest
the decomposition of the complexified Lorentz algebra into two
commuting complex $\mathfrak{sl}_2$ factors,
\begin{equation}
    \mathfrak{so}(3,1)_{\mathbb C}
    \simeq
    \mathfrak{so}(4,\mathbb C)
    \simeq
    \mathfrak{sl}_2(\mathbb C)_L
    \oplus
    \mathfrak{sl}_2(\mathbb C)_R
    \subset
    \mathfrak{so}(3,2)_{\mathbb C}
    \simeq
    \mathfrak{so}(5,\mathbb C).
\end{equation}
Each factor has three complex generators, which we denote by
$L_{0,\pm1}$ and $\bar L_{0,\pm1}$.  The mode construction below
uses the corresponding complex $\mathfrak{sl}_2$ weight
decomposition, with Cartan labels $m$ and $\bar m$. 

The complexification does not change the
physical Lorentzian AdS$_4$ background.  Lorentzian reality defines
an antilinear involution that exchanges the dotted and undotted
$\mathfrak{sl}_2(\mathbb C)$ factors. It has real algebra as
$\mathfrak{so}(3,1)$\footnote{  This is distinct from imposing separate
reality conditions on the two factors,
\begin{equation}
    \mathfrak{sl}_2(\mathbb R)_L
    \oplus
    \mathfrak{sl}_2(\mathbb R)_R
    \simeq
    \mathfrak{so}(2,2),
\end{equation}
which selects the split real form of the six-generator Lorentz
subalgebra.}.  We therefore derive the mode algebra in the
complexified representation and impose the appropriate Lorentzian
reality conditions on physical fields separately (see also
Appendix~\ref{app:spinors-real-forms}).

 

\subsection{The AdS spinor-helicity representation}
\label{subsec:ads_spinor_helicity}

We introduce independent two-component spinors
$\lambda_\alpha$ and $\bar\lambda_{\dot\alpha}$.
The Lorentz generators can be written in these spinors as
\begin{align}
    \mathcal M_{\alpha\beta}
    &=
    \lambda_\alpha
    \frac{\partial}{\partial\lambda^\beta}
    +
    \lambda_\beta
    \frac{\partial}{\partial\lambda^\alpha},
    \label{eq:ads_M}\\
    \bar{\mathcal M}_{\dot\alpha\dot\beta}
    &=
    \bar\lambda_{\dot\alpha}
    \frac{\partial}{\partial\bar\lambda^{\dot\beta}}
    +
    \bar\lambda_{\dot\beta}
    \frac{\partial}{\partial\bar\lambda^{\dot\alpha}}.
    \label{eq:ads_Mbar}
\end{align}
The four remaining generators, the AdS transvections, are
\begin{equation}
    \mathcal P_{\alpha\dot\alpha}
    =
    \lambda_\alpha\bar\lambda_{\dot\alpha}
    +
    \Lambda
    \frac{\partial}{\partial\lambda^\alpha}
    \frac{\partial}{\partial\bar\lambda^{\dot\alpha}},
    \qquad
    \Lambda=-\ell^{-2}.
    \label{eq:ads_transvection}
\end{equation}
Together these operators furnish the AdS spinor-helicity
representation
\cite{NagarajPonomarev:2019,NagarajPonomarev:2020}.
The curvature dependence is entirely in the second term of
eq.~\eqref{eq:ads_transvection}, which disappears in the flat limit. Using these spinors, one can write the helicity operator as

\begin{equation}
    \mathfrak h
    =
    \frac12
    \left(
        \bar\lambda_{\dot\alpha}
        \frac{\partial}{\partial\bar\lambda_{\dot\alpha}}
        -
        \lambda_\alpha
        \frac{\partial}{\partial\lambda_\alpha}
    \right).
    \label{eq:helicity_operator}
\end{equation}
A helicity-$\sigma$ wavefunction obeys
\begin{equation}
    \mathfrak h\Psi_\sigma=\sigma\Psi_\sigma,
    \label{eq:helicity_eigenvalue}
\end{equation}
or, equivalently,
\begin{equation}
    \Psi_\sigma(t\lambda,t^{-1}\bar\lambda)
    =
    t^{-2\sigma}\Psi_\sigma(\lambda,\bar\lambda).
    \label{eq:little_group_ads}
\end{equation}
For gravitons, $\sigma=\pm2$.

\subsection{Projective spinors and the Lorentz action}
\label{subsec:projective_spinors}

We can separate the projective coordinates from the overall scale
\begin{align}
    \lambda_\alpha
    &=
    a\,q_\alpha(z),
    & q_\alpha(z)&=(1,z),
    \label{eq:lambda_projective}\\
    \bar\lambda_{\dot\alpha}
    &=
    b\,\bar q_{\dot\alpha}(\bar z),
    & \bar q_{\dot\alpha}(\bar z)&=(1,\bar z).
    \label{eq:lambdabar_projective}
\end{align}
The coordinates $z$ and $\bar z$ are independent in the
complexified calculation. Define
\begin{equation}
    \omega=ab,
    \qquad
    \rho=\frac{a}{b}.
    \label{eq:omega_rho}
\end{equation}
Then the Euler operators of these variables are
\begin{equation}
    a\partial_a
    =
    \omega\partial_\omega+\rho\partial_\rho,
    \qquad
    b\partial_b
    =
    \omega\partial_\omega-\rho\partial_\rho.
    \label{eq:euler_ab}
\end{equation}
Equation~\eqref{eq:little_group_ads} fixes the relative-scale
dependence:
\begin{equation}
    \rho\partial_\rho\Psi_\sigma
    =
    -\sigma\Psi_\sigma.
    \label{eq:rho_helicity}
\end{equation}
Thus $\rho$ carries the little-group weight, while $\omega$
is the remaining common scale. We may choose $\rho=1$ when
labeling the solutions by $(\omega,z,\bar z)$. The general
$\rho$ dependence is restored by multiplication by $\rho^{-\sigma}$.

Choosing global generators $L_r$ and $\bar L_{\bar r}$ with
$r,\bar r=-1,0,1$, the Lorentz action becomes
\begin{align}
    L_r
    &=
    -z^{r+1}\partial_z
    +
    \frac{r+1}{2}z^r
    \left(\omega\partial_\omega-\sigma\right),
    \label{eq:Lr_before_mellin}\\
    \bar L_{\bar r}
    &=
    -\bar z^{\bar r+1}\partial_{\bar z}
    +
    \frac{\bar r+1}{2}\bar z^{\bar r}
    \left(\omega\partial_\omega+\sigma\right).
    \label{eq:Lbar_before_mellin}
\end{align}
These expressions follow from
Eqs.~\eqref{eq:ads_M}-\eqref{eq:ads_Mbar} and
\eqref{eq:euler_ab}.

\subsection{Mellin transformation and Lorentz weights}
\label{subsec:mellin_homogeneous_basis}
\label{subsec:mellin_preserves_eom}

Let $\Psi_\sigma(x;\omega,z,\bar z)$ denote an on-shell
wavefunction, with tensor indices suppressed.
Define its Mellin transform by
\begin{equation}
    \widetilde\Psi_{\Delta,\sigma}(x;z,\bar z)
    =
    \int_0^\infty d\omega\,
    \omega^{\Delta-1}
    \Psi_\sigma(x;\omega,z,\bar z).
    \label{eq:ads_mellin_transform}
\end{equation}
The transform is defined initially in a convergence domain, and then continued
analytically or distributionally in $\Delta$. Because the linearized Einstein and curvature operators act only
on spacetime variables, Mellin transformation preserves the
on-shell relations:
\begin{align}
    \mathcal E_{\mu\nu}
    [\widetilde h^{(\sigma)}_\Delta]
    &=
    \int_0^\infty d\omega\,
    \omega^{\Delta-1}
    \mathcal E_{\mu\nu}[h^{(\sigma)}(\omega)]
    =0,
    \label{eq:mellin_einstein}\\
    C^{(1)}[\widetilde h^{(\sigma)}_\Delta]
    &=
    \int_0^\infty d\omega\,
    \omega^{\Delta-1}C^{(1)}[h^{(\sigma)}(\omega)]
    =
    \widetilde C^{(\sigma)}_\Delta.
    \label{eq:mellin_weyl}
\end{align}
Here $C^{(1)}$ denotes the linearized Weyl operator. To determine the Lorentz weights, integrate the scale derivative
by parts:
\begin{equation}
    \int_0^\infty d\omega\,
    \omega^{\Delta-1}\omega\partial_\omega\Psi_\sigma
    =
    \left[\omega^\Delta\Psi_\sigma\right]_0^\infty
    -
    \Delta\,\widetilde\Psi_{\Delta,\sigma}.
    \label{eq:mellin_euler}
\end{equation}
When the endpoint term vanishes, and subsequently under the
chosen continuation, the scale Euler operator is represented
by $-\Delta$. Equations~\eqref{eq:Lr_before_mellin} and
\eqref{eq:Lbar_before_mellin} therefore become
\begin{align}
    L_r
    &=
    -z^{r+1}\partial_z
    -
    \frac{r+1}{2}(\Delta+\sigma)z^r,
    \label{eq:Lr_after_mellin}\\
    \bar L_{\bar r}
    &=
    -\bar z^{\bar r+1}\partial_{\bar z}
    -
    \frac{\bar r+1}{2}(\Delta-\sigma)\bar z^{\bar r}.
    \label{eq:Lbar_after_mellin}
\end{align}
Comparing with the projective action
$L_r=-z^{r+1}\partial_z-(r+1)h z^r$
and its barred counterpart gives
\begin{equation}
        h=\frac{\Delta+\sigma}{2},
        \qquad
        \bar h=\frac{\Delta-\sigma}{2}.
    \label{eq:weights_general}
\end{equation}
In particular,
\begin{equation}
    h+\bar h=\Delta,
    \qquad
    h-\bar h=\sigma.
    \label{eq:weight_relations}
\end{equation}
For the two graviton helicities,
\begin{align}
    (\bar h,h)_{\sigma=+2}
    &=
    \left(\frac{\Delta-2}{2},\frac{\Delta+2}{2}\right),
    \label{eq:graviton_weights_plus}\\
    (\bar h,h)_{\sigma=-2}
    &=
    \left(\frac{\Delta+2}{2},\frac{\Delta-2}{2}\right).
    \label{eq:graviton_weights_minus}
\end{align}

These weights have the same form as those of the flat-space
conformal basis \cite{PasterskiShao:2017}, since they follow from
the Lorentz action and little-group homogeneity. Here, however,
the underlying fields solve the AdS equations, and $\Delta$
labels spinor-scale homogeneity rather than the dimension of a
three-dimensional boundary primary. The two Lorentz factors preserve each fixed-$\Delta$ sector. But the AdS transvection generator Eq.~\eqref{eq:ads_transvection} does not preserve it. The multiplication and
derivative terms in Eq.~\eqref{eq:ads_transvection} have
opposite common-scale degrees and connect Mellin sectors differing
by one unit. 


\section{Laurent completion and the full AdS wedge}
\label{subsec:laurent_completion_ads_wedge}

\subsection{Laurent expansion of the bulk graviton}
\label{subsec:ads_laurent_modes}

Let
\begin{equation}
    \mathcal H^{p,+}_{\mu\nu}(x;z,\bar z)
    =
    \left.
        \int_0^\infty d\omega\,\omega^{\Delta-1}
        h^{(+)}_{\mu\nu}(x;\omega,z,\bar z)
    \right|_{\Delta=4-2p},
    \qquad
    (\bar h,h)=(1-p,3-p).
    \label{eq:bulk_mellin_graviton_p}
\end{equation}
Here we are parameterizing $\Delta=4-2p$. The resulting field satisfies the Einstein eq
$\mathcal E_x[\mathcal H^{p,+}]=0$.
The action of the Lorentz generator acting on $\mathcal H^{p,+}$ can be evaluated as \footnote{By Lorentz covariance, the commutator action on the bulk fields is represented by the negative of the differential action on the spinor-helicity labels in Eq.~\eqref{eq:Lr_after_mellin}-\eqref{eq:Lbar_after_mellin}.}
\begin{align}
    [L_r,\mathcal H^{p,+}]
    &=
    \left[
        z^{r+1}\partial_z+(r+1)(3-p)z^r
    \right]\mathcal H^{p,+},
    \\
    [\bar L_r,\mathcal H^{p,+}]
    &=
    \left[
        \bar z^{r+1}\partial_{\bar z}
        +(r+1)(1-p)\bar z^r
    \right]\mathcal H^{p,+},
    \qquad r=-1,0,1.
    \label{eq:bulk_primary_active_action}
\end{align}

Define mode labels by
$[L_0,w^p_{\bar m,m}]=-m w^p_{\bar m,m}$ and
$[\bar L_0,w^p_{\bar m,m}]=-\bar m w^p_{\bar m,m}$.
Exactly as for a primary-field Laurent expansion, these
eigenvalues fix the powers:
\begin{equation}
    \mathcal H^{p,+}_{\mu\nu}(x;z,\bar z)
    =
    \sum_{\bar m,m}
    \bar z^{\,p-1-\bar m}\,
    z^{\,p-3-m}\,
    \frac{w^p_{\bar m,m;\mu\nu}(x)}
         {N_{p,\bar m}}.
    \label{eq:direct_graviton_mode_expansion}
\end{equation}
Here $N_{p,\bar m}$ is an $x$-independent normalization. The sums run over
labels for which the Laurent powers are integers.

The modes can be obtained directly from the bulk field as
\begin{equation}
    w^p_{\bar m,m;\mu\nu}(x)
    =
    N_{p,\bar m}
    \oint_{z=0}\frac{dz}{2\pi i}\,z^{m+2-p}
    \oint_{\bar z=0}\frac{d\bar z}{2\pi i}\,
        \bar z^{\bar m-p}
        \mathcal H^{p,+}_{\mu\nu}(x;z,\bar z).
    \label{eq:w_directly_from_graviton}
\end{equation}
The corresponding modes also satisfy Einstein's equations.
\begin{equation}
    \mathcal E_x[w^p_{\bar m,m}]=0.
    \label{eq:graviton_modes_on_shell}
\end{equation}
Thus every mode is an on-shell metric
perturbation.
\subsection*{Normalized modes}
Now let's restrict the chosen soft sector to the finite
polynomial basis,
\begin{equation}
    p=1,\frac32,2,\ldots,
    \qquad
    \partial_{\bar z}^{\,2p-1}\mathcal H^{p,+}=0.
    \label{eq:finite_soft_seed_condition}
\end{equation}
Then $1-p\leq\bar m\leq p-1$. To express this finite multiplet
in the standard spin-$(p-1)$ basis, the raising relation
requires
\begin{equation}
    \frac{N_{p,\bar m+1}}{N_{p,\bar m}}
    =
    -\frac{p+\bar m}{p-1-\bar m}.
\end{equation}
Its solution is
\begin{equation}
    N_{p,\bar m}
    =
    (-1)^{p-1-\bar m}
    \Gamma(p+\bar m)\Gamma(p-\bar m).
    \label{eq:finite_mode_normalization}
\end{equation}
The factorial factors are the familiar soft-mode normalization
\cite{Strominger:2021}. The phase follows from eq.~\eqref{eq:bulk_primary_active_action}. Using the action of $L_r, \bar{L}_r$ on $\mathcal{H}^p$, we can find the action of these Lorentz generators on the modes as
\begin{equation}
    [\bar L_r,w^p_{\bar m,m}]
    =
    (-rp-\bar m)
    \frac{N_{p,\bar m}}{N_{p,\bar m+r}}\,
    w^p_{\bar m+r,m}.
\end{equation}
The normalization ratios are
\begin{equation}
    \frac{N_{p,\bar m}}{N_{p,\bar m+1}}
    =
    -\frac{p-1-\bar m}{p+\bar m},
    \qquad
    \frac{N_{p,\bar m}}{N_{p,\bar m-1}}
    =
    -\frac{p-1+\bar m}{p-\bar m}.
\end{equation}
Substituting these ratios, and treating the finite endpoints
by polynomial truncation, yields
\begin{align}
    [\bar L_r,w^p_{\bar m,m}]
    &=
    [r(p-1)-\bar m]\,
    w^p_{\bar m+r,m},
    \label{eq:w_full_bar_action}\\
    [L_r,w^p_{\bar m,m}]
    &=
    [r(2-p)-m]\,
    w^p_{\bar m,m+r},
    \label{eq:w_full_unbar_action}
\end{align}
for $r=-1,0,1$. Comparing with the standard mode convention
$[L_r,w_m]=[r(h_w-1)-m]w_{m+r}$, we therefore assign
\begin{equation}
(\bar h_w,h_w)=(p,3-p).
\end{equation}
\begin{equation}
    [\bar L_0,w^p_{\bar m,m}]
    =-\bar m\,w^p_{\bar m,m},
    \qquad
    \bar{\mathcal C}\,w^p_{\bar m,m}
    =p(p-1)w^p_{\bar m,m}.
\end{equation}


\paragraph{ Laurent completion:-}
We make one \emph{Laurent-completion assumption}: the normalized
bulk-mode construction admits an AdS-equivariant extension to
ordinary, integer-power Laurent modes, retaining every compatible
Mellin sector, not just the finite polynomial submodules.
The word ``normalized'' here does not mean that the Gamma factors in
Eq.~\eqref{eq:finite_mode_normalization} are analytically continued
through their poles. We assume that the modes
themselves admit an AdS-equivariant continuation, with their
normalization fixed by the Lorentz action
\begin{align}
    [\bar L_r,w^p_{\bar m,m}]
    &=
    [r(p-1)-\bar m]\,
    w^p_{\bar m+r,m},
    \\
    [L_r,w^p_{\bar m,m}]
    &=
    [r(2-p)-m]\,
    w^p_{\bar m,m+r}.
    \label{eq:completed-Lorentz-normalization}
\end{align}
One lowest-weight state at each value of $p$ fixes the overall
normalization of that module.

This prescription is also the one naturally realized in the
CFT$_3$ construction of Strominger and Wei
\cite{StromingerWei:2026}. Their lowest-weight states are fixed
directly by the Fourier modes of the ANEC operator,
\begin{equation}
    \mathcal E_k
    =
    -i^k\,
    w^{\frac{3+k}{2}}_{-\frac{k+1}{2},\,\frac{k-1}{2}}.
\end{equation}
Setting $k=2p-3$ gives
\begin{equation}
    \mathcal E_{2p-3}
    =
    -i^{\,2p-3}\,
    w^p_{1-p,p-2},
    \label{eq:ANEC-lowest-weight-normalization}
\end{equation}
which supplies a normalized lowest-weight state for every
$p\in\frac12\mathbb Z$. The remaining modes are then normalized by
the $SO(3,2)$ action. In particular, their Lorentz action is
\begin{align}
    [Q(\bar L_1),w^p_{\bar m,m}]
    &=
    (p-1-\bar m)\,
    w^p_{\bar m+1,m},
    \\
    [Q(L_1),w^p_{\bar m,m}]
    &=
    (2-p-m)\,
    w^p_{\bar m,m+1},
\end{align}
which agrees with
Eq.~\eqref{eq:completed-Lorentz-normalization}.
Thus the boundary construction uses the same canonical
lowest-weight-module normalization.

The lattice follows from the Laurent powers and the lowering
zeros. The original graviton expansion contains the factors
$\bar z^{p-1-\bar m}z^{p-3-m}$, so ordinary Laurent modes require
\begin{equation}
    p-1-\bar m\in\mathbb Z,
    \qquad
    p-3-m\in\mathbb Z.
    \label{eq:integer_laurent_powers}
\end{equation}
The two lowering coefficients vanish at
$\bar m=1-p$ and $m=p-2$. At this corner the Laurent factor is
$\bar z^{2p-2}z^{-1}$. Hence
\begin{equation}
    2p-2\in\mathbb Z
    \qquad\Longleftrightarrow\qquad
    p\in\tfrac12\mathbb Z.
\end{equation}
Polynomial shortening additionally required $2p-2\geq0$.
Laurent expansion permits negative powers, so this restriction
is absent in the completion. Combining the half-integer
$p$-lattice with Eq.~\eqref{eq:integer_laurent_powers} gives ($m,\bar{m}$ are half integer)
\begin{equation}
    p\pm\bar m\in\mathbb Z,
    \qquad
    p\pm m\in\mathbb Z.
    \label{eq:mode-lattice}
\end{equation}
At fixed $p$, the $\bar L_1$ acts as
\begin{equation}
    [\bar L_1,w^p_{\bar m,m}]
    =
    (p-1-\bar m)\,w^p_{\bar m+1,m}.
\end{equation}
Starting from the lowest allowed index $\bar m=1-p$, write
$\bar m=1-p+a$, with $a\in\mathbb Z_{\geq0}$. The raising
coefficient is then $2p-2-a$. For $p\geq1$, it vanishes at
$a=2p-2$, terminating the descendants of the lowest mode at
$\bar m=p-1$. For $p<1$, however, it never vanishes for any
$a\geq0$. Repeated raising therefore produces the infinite tower
\begin{equation}
    w^p_{1-p,m}
    \;\xrightarrow{\;\bar L_1\;}\;
    w^p_{2-p,m}
    \;\xrightarrow{\;\bar L_1\;}\;
    w^p_{3-p,m}
    \;\xrightarrow{\;\bar L_1\;}\;\cdots .
\end{equation}
Thus the $p<1$ sectors included in the Laurent completion have no truncation. 

The AdS transvections act between these fixed-$p$ sectors.
They map a mode at level $p$ to a linear combination of modes
at levels $p-\tfrac12$ and $p+\tfrac12$, with corresponding
half-integer shifts of $\bar m$ and $m$. 

\section{Dirichlet boundary condition and helicity pairing}
\label{sec:dirichlet-global}
\subsection{Boundary data and helicity pairing}
\label{sec:helicity-pairing}

Let $\mathcal E_{ij}$ and $\mathcal B_{ij}$ denote the finite
electric and magnetic boundary data of the linearized Weyl tensor.
The organization of asymptotically AdS gravitational data in terms of
the boundary stress tensor and electric Weyl tensor is standard in
holographic renormalization
\cite{BalasubramanianKraus:1999,deHaroSkenderisSolodukhin:2001,
BianchiFreedmanSkenderis:2001,Skenderis:2002,AshtekarDas:1999}.
In four bulk dimensions, the complementary magnetic data are related
to the Cotton tensor of the boundary metric and play a natural role
in gravitational electric--magnetic duality
\cite{LeighPetkou:2007,deHaroPetkou:2007,Bakas:2009}.
At the linearized level,
\begin{equation}
    \mathcal E_{ij}
    \propto
    \delta\langle T_{ij}\rangle,
    \qquad
    \mathcal B_{ij}
    \propto
    \delta C_{ij}[g_{(0)}].
    \label{eq:EB-boundary}
\end{equation}

Source-free Dirichlet boundary conditions fix the boundary metric \footnote{The converse requires additional requirements. Vanishing of  magnetic data constrains
the boundary Cotton tensor, but does not by itself fix the
boundary metric representative. 
We therefore use $\delta g_{(0)ij}=0$ as the boundary condition
and $\mathcal B_{ij}=0$ to determine its implication for
helicity pairing.},
\begin{equation}
    \delta g_{(0)ij}=0,
\end{equation}
and therefore imply
\begin{equation}
    \mathcal B_{ij}=0.
    \label{eq:dirichlet-source}
\end{equation}
More general choices in which the boundary metric is allowed to
fluctuate lead to Neumann or mixed gravitational boundary conditions
\cite{CompereMarolf:2008}.


With Lorentzian Hodge duality ${}^\star{}^\star=-1$, define
\begin{equation}
    C^{(\pm)}=\frac12\bigl(C\mp i\,{}^\star C\bigr),
    \qquad
    {}^\star C^{(\pm)}=\pm i C^{(\pm)}.
\end{equation}
Taking the magnetic data to be the electric projection of
${}^\star C$, the two chiral sectors obey
\begin{equation}
    \mathcal B^{(+)}_{ij}=i\mathcal E^{(+)}_{ij},
    \qquad
    \mathcal B^{(-)}_{ij}=-i\mathcal E^{(-)}_{ij}.
\end{equation}
For their sum, the vanishing of the magnetic data is thus
equivalent to
\begin{equation}
    \mathcal E^{(+)}_{ij}=\mathcal E^{(-)}_{ij}.
    \label{eq:equal-electric-data}
\end{equation}
This equality of the electric contributions is the direct consequence of the Dirichlet boundary condition. We now determine the form of the matching map $\mathcal R_D$ between the
two chiral modes. Writing $\mathcal E_\pm$ for
their boundary electric-data maps, its defining condition is
\begin{equation}
    \mathcal E_-\bigl(\mathcal R_D w\bigr)
    =\mathcal E_+(w).
    \label{eq:RD-electric-matching}
\end{equation}
The positive and negative helicity towers carry weights
\begin{equation}
    (\bar h,h)_+=(p,3-p),
    \qquad
    (\bar h,h)_-=(3-q,q).
\end{equation}
 The quadratic Casimir of $\mathfrak{sl}_2$ module of
weight $h$ is  $h(h-1)$. And the difference
$\mathcal C_{\bar L}-\mathcal C_L$ is
\begin{align}
    (\mathcal C_{\bar L}-\mathcal C_L)_+
    &=p(p-1)-(3-p)(2-p)=4p-6,
    \\
    (\mathcal C_{\bar L}-\mathcal C_L)_-
    &=(3-q)(2-q)-q(q-1)=6-4q.
\end{align}
A nonzero intertwiner must preserve these eigenvalues and
therefore pairs
\begin{equation}
    q=3-p.
    \label{eq:q-three-minus-p}
\end{equation}
At these levels both Lorentz weights agree. Intertwining the
two Cartan generators then preserves the indices
$(\bar m,m)$. 
\begin{equation}
    \mathcal R_D w^p_{\bar m,m}
    =
    \rho_{p;\bar m,m}\,
    \bar w^{\,3-p}_{\bar m,m}.
    \label{eq:RD-mode-map}
\end{equation}
Here the bar on $\bar w$ labels the opposite-helicity family.
The opposite-helicity wedge conditions,
$m+q\geq1$ and $\bar m-q\geq-2$, become
$m-p+2\geq0$ and $\bar m+p-1\geq0$, precisely the original
wedge.

To determine the dependence of the coefficient on the mode
indices, introduce the nonnegative integers
\begin{equation}
    a=\bar m+p-1,
    \qquad
    c=m-p+2.
    \label{eq:wedge-distances}
\end{equation}
After the identification $q=3-p$, the Lorentz lowering
generators act with the same coefficients in both modules.
\begin{equation}
    \bar L_{-1}\cdot w^p_{\bar m,m}
    =-a\,w^p_{\bar m-1,m},
    \qquad
    L_{-1}\cdot w^p_{\bar m,m}
    =-c\,w^p_{\bar m,m-1}.
\end{equation}
Consequently, equivariance of $\mathcal R_D$ gives
\begin{align}
    \rho_{p;\bar m,m}
    &=\rho_{p;\bar m-1,m},
    &&a>0,
    \\
    \rho_{p;\bar m,m}
    &=\rho_{p;\bar m,m-1},
    &&c>0.
\end{align}
Every point in the wedge can be lowered to the corner
$(\bar m,m)=(1-p,p-2)$ without encountering a vanishing
coefficient before the corresponding boundary is reached.
Thus the relative coefficient depends only on the level,
\begin{equation}
    \rho_{p;\bar m,m}=\rho_p,
    \qquad
    W^p_{\bar m,m}
    =
    w^p_{\bar m,m}
    +
    \rho_p\,\bar w^{\,3-p}_{\bar m,m}.
    \label{eq:W-rho}
\end{equation}
Its boundary normalization is fixed by
\begin{equation}
    \mathcal E_+\bigl(w^p_{\bar m,m}\bigr)
    =
    \rho_p\,
    \mathcal E_-\bigl(\bar w^{\,3-p}_{\bar m,m}\bigr).
    \label{eq:rho-electric-normalization}
\end{equation}

This argument fixes the form of any Lorentz-equivariant
electric-data pairing on the specified modules (see Appendix \ref {sec:metric-dirichlet-pairing} for metric realization of the pairing). 
Lorentz covariance makes $\rho_p$ independent of the mode
indices and compatibility with the AdS transvections will
relate its values at neighboring levels. This is what we do next.

\subsection{Transvections and the global soft seed}
\label{sec:global-action}

Using the normalization
$H_{\bar r,r}=-\frac12\mathcal P_{\bar r,r}$, with
$r,\bar r=\pm\frac12$, we define
\begin{equation}
    A_p(\bar r,\bar m)
    =
    \bar r(p-1)-\frac{\bar m}{2},
    \qquad
    B_p(r,m)
    =
    r(p-2)+\frac{m}{2}.
    \label{eq:AB}
\end{equation}
The action of transvection on the modes can be evaluated as (see Appendix \ref{app:adstranvection} for a complete derivation)
\begin{align}
    [H_{\bar r,r},w^p_{\bar m,m}]
    &=
    A_p\,w^{p-\frac12}_{\bar m+\bar r,m+r}
    -
    \Lambda B_p\,w^{p+\frac12}_{\bar m+\bar r,m+r}.
    \label{eq:H-action-SD}
\end{align}
Exchanging the two Lorentz factors gives the corresponding
opposite-helicity action
\begin{align}
    [H_{\bar r,r},\bar w^{\,3-p}_{\bar m,m}]
    &=
    -B_p\,\bar w^{\,\frac52-p}_{\bar m+\bar r,m+r}
    +
    \Lambda A_p\,
    \bar w^{\,\frac72-p}_{\bar m+\bar r,m+r}.
    \label{eq:H-action-matched-ASD}
\end{align}
For the AdS Dirichlet combination \eqref{eq:W-rho} to be invariant under these
transvections, its coefficients must obey
\begin{equation}
    \rho_{p-\frac12}=\Lambda\rho_p,
    \qquad
    \rho_{p+\frac12}=\Lambda^{-1}\rho_p.
    \label{eq:rho-recursion}
\end{equation}
At $p=\frac32$,
$\rho_{3/2}=1$, fixes
\begin{equation}
    W^p_{\bar m,m}
    =
    w^p_{\bar m,m}
    +
    \Lambda^{\,3-2p}\bar w^{\,3-p}_{\bar m,m}.
    \label{eq:Dirichlet-mode}
\end{equation}
The exponent is an integer on the half-integer lattice.  This relative
normalization is specific to the chiral basis and transvection
conventions that we have chosen here.

Then, the Dirichlet graviton modes carry the action of the generator as
\begin{align}
    [\bar L_k,W^p_{\bar m,m}]
    &=
    [k(p-1)-\bar m]\,W^p_{\bar m+k,m},
    \label{eq:Dirichlet-barL}\\
    [L_k,W^p_{\bar m,m}]
    &=
    [k(2-p)-m]\,W^p_{\bar m,m+k},
    \label{eq:Dirichlet-L}\\
    [H_{\bar r,r},W^p_{\bar m,m}]
    &=
    A_p\,W^{p-\frac12}_{\bar m+\bar r,m+r}
    -
    \Lambda B_p\,W^{p+\frac12}_{\bar m+\bar r,m+r},
    \label{eq:H-action-W}
\end{align}
where $k=-1,0,1$ and $r,\bar r=\pm\frac12$.  These are the same representation matrices as in
the chiral wedge, so the wedge remains invariant.

We identify the ten distinguished modes (AdS isometry generators) along with their normalization by
\begin{equation}
    \bar L_k=W^2_{k,0},
    \qquad
    L_k=\Lambda^{-1}W^1_{0,k},
    \qquad
    H_{\bar r,r}=W^{3/2}_{\bar r,r}.
    \label{eq:global-identification}
\end{equation}
For the bracket constructed below, requiring these modes to act by
\eqref{eq:Dirichlet-barL}-\eqref{eq:H-action-W} is the
\emph{global soft seed}. The geometric action obeys
\begin{align}
    [L_k,L_l]&=(k-l)L_{k+l},
    &
    [\bar L_k,\bar L_l]&=(k-l)\bar L_{k+l},
    &
    [L_k,\bar L_l]&=0,
    \\
    [L_k,H_{\bar r,r}]
    &=
    \left(\frac{k}{2}-r\right)H_{\bar r,k+r},
    &
    [\bar L_k,H_{\bar r,r}]
    &=
    \left(\frac{k}{2}-\bar r\right)H_{k+\bar r,r},
\end{align}
and
\begin{equation}
    [H_{\bar r,r},H_{\bar s,s}]
    =
    \frac{\Lambda}{2}
    \left[
        (\bar r-\bar s)L_{r+s}
        +(r-s)\bar L_{\bar r+\bar s}
    \right].
    \label{eq:HH-global}
\end{equation}
Together these furnish the AdS isometry algebra.

\section{Bootstrap of the classical linear soft algebra}
\label{sec:bootstrap}
The Lie Brackets for the global modes are written in eq. \eqref{eq:HH-global}. The brackets are bilinear and antisymmetric. We wish to generalize the global algebra to the whole AdS modes tower.  We require linear closure,
with no central or nonlinear terms. We also require the global soft
seed \eqref{eq:global-identification} algebra. The linear closure can be motivated from the charge algebra perspective as well \cite{DiGiacomo:2026}. 

\subsection{The two output levels}
\label{sec:two-levels}

Jacobi with the two Cartan generators requires addition of mode numbers\footnote{The mode indices are fixed by the two Cartan generators.
For $X=W^p_{\bar m,m}$ and $Y=W^q_{\bar n,n}$, their known
global action gives
\begin{equation}
    [L_0,X]=-mX,
    \qquad
    [\bar L_0,X]=-\bar mX,
\end{equation}
and analogously for $Y$. Jacobi therefore implies
\begin{align}
    [L_0,[X,Y]]
    &=
    [[L_0,X],Y]+[X,[L_0,Y]]
    =-(m+n)[X,Y],
    \\
    [\bar L_0,[X,Y]]
    &=
    [[\bar L_0,X],Y]+[X,[\bar L_0,Y]]
    =-(\bar m+\bar n)[X,Y].
\end{align}
Then the linear closure in the simultaneous Cartan
eigenbasis requires
\begin{equation}
    [W^p_{\bar m,m},W^q_{\bar n,n}]
    =
    \sum_s C^s_{p,q}(\bar m,m;\bar n,n)\,
    W^s_{\bar m+\bar n,m+n}.
    \label{eq:cartan-mode-addition}
\end{equation}
Only the output level $s$ remains undetermined.}
\begin{equation}
    [W^p_{\bar m,m},W^q_{\bar n,n}]
    =
    \sum_s C^s\,W^s_{\bar m+\bar n,m+n}.
    \label{eq:cartan-mode-addition}
\end{equation}
For the two inputs, introduce the nonnegative distances
\begin{equation}
    a=\bar m+p-1,\qquad b=\bar n+q-1,
    \qquad
    c=m-p+2,\qquad d=n-q+2.
\end{equation}
The barred lowering operator annihilates the inputs after $a+1$ and
$b+1$ steps.  Since it acts as a derivation, it annihilates
their bracket after $a+b+1$ steps.  An output at level $s$
has barred distance
\begin{equation}
    a_{\mathrm{out}}=a+b+s-p-q+1,
\end{equation}
and hence a nonzero coefficient requires $s\leq p+q-1$.
The unbarred lowering operator similarly gives
\begin{equation}
    c_{\mathrm{out}}=c+d+p+q-s-2,
    \qquad
    s\geq p+q-2.
\end{equation}
Finally, the mode lattice implies $s-p-q\in\mathbb Z$.  Therefore
\begin{equation}
s=p+q-2\quad\text{or}\quad s=p+q-1.
    \label{eq:two-output-levels}
\end{equation}

\subsection{Lowering covariance and propagation of the seed}
\label{sec:transvection-bootstrap}
To ease up the notation, we define a mode $V^p_{a,c}$ as
\begin{equation}
    V^p_{a,c}
    \equiv
    W^p_{1-p+a,p-2+c}.
\end{equation}
The general bracket becomes\footnote{The first-step coefficients are fixed directly by two global
transvections. Since
\begin{equation}
    V^{3/2}_{1,0}=H_{\frac12,-\frac12},
    \qquad
    V^{3/2}_{0,1}=H_{-\frac12,\frac12},
\end{equation}
their independently known action on the corner $V^q_{0,0}$
gives
\begin{align}
    [V^{3/2}_{1,0},V^q_{0,0}]
    &=
    (q-1)V^{q-\frac12}_{0,0},
    \\
    [V^{3/2}_{0,1},V^q_{0,0}]
    &=
    -\Lambda(q-2)V^{q+\frac12}_{0,0}.
\end{align}
The first commutator isolates $F_{3/2,q}(1,0)$, and the
second isolates $G_{3/2,q}(1,0)$. Hence
\begin{equation}
    \alpha_{\frac32,q}=q-1,
    \qquad
    \beta_{\frac32,q}=-\Lambda(q-2).
    \label{eq:first-step-seed}
\end{equation}}
\begin{align}
    [V^p_{a,c},V^q_{b,d}]
    &=
    F_{p,q}(a,b;c,d)\,
    V^{p+q-2}_{a+b-1,c+d}
    +
    G_{p,q}(a,b;c,d)\,
    V^{p+q-1}_{a+b,c+d-1}.
    \label{eq:FG-general}
\end{align}
A term is absent whenever its output has a negative distance
index. Thus $F_{p,q}(0,0;c,d)=0$ and
$G_{p,q}(a,b;0,0)=0$.

For the first term, Jacobi with $L_{-1}$ gives
\begin{equation}
    (c+d)F_{p,q}(a,b;c,d)
    =
    cF_{p,q}(a,b;c-1,d)
    +
    dF_{p,q}(a,b;c,d-1).
\end{equation}
Induction on $c+d$ makes $F$ independent of $c,d$.
Barred lowering then gives
\begin{equation}
    (a+b-1)F_{p,q}(a,b)
    =
    aF_{p,q}(a-1,b)
    +
    bF_{p,q}(a,b-1),
    \qquad
    F_{p,q}(0,0)=0.
    \label{eq:F-lowering}
\end{equation}
Thus $F$ is determined by its two first-step values.
The analogous argument makes $G$ independent of $a,b$ and linear
in $c,d$.  Including antisymmetry, the general result is
\begin{equation}
    F_{p,q}(a,b)
    =
    \alpha_{p,q}a-\alpha_{q,p}b,
    \qquad
    G_{p,q}(c,d)
    =
    \beta_{p,q}c-\beta_{q,p}d.
    \label{eq:first-step-coefficients}
\end{equation}

Now use the lowering transvection
$H_-\equiv H_{-\frac12,-\frac12}$.  Equation
\eqref{eq:H-action-W} becomes
\begin{equation}
    [H_-,V^p_{a,c}]
    =
    -\frac{a}{2}\,V^{p-\frac12}_{a-1,c}
    -
    \frac{\Lambda c}{2}\,V^{p+\frac12}_{a,c-1}.
    \label{eq:Hminus-distance}
\end{equation}
In the Jacobi identity
\begin{equation}
    [H_-,[X,Y]]
    =
    [[H_-,X],Y]+[X,[H_-,Y]],
    \label{eq:Hminus-Jacobi}
\end{equation}
the lowest output level, $p+q-\frac52$, receives only the first
term.  Its coefficient gives
\begin{equation}
    (a+b-1)F_{p,q}(a,b)
    =
    aF_{p-\frac12,q}(a-1,b)
    +
    bF_{p,q-\frac12}(a,b-1).
    \label{eq:F-transvection-recursion}
\end{equation}
Setting $(a,b)=(2,0)$ yields
\begin{equation}
    \alpha_{p,q}=\alpha_{p-\frac12,q}.
    \label{eq:alpha-propagation}
\end{equation}
Likewise, the highest output level, $p+q-\frac12$, receives only
the second term.  Cancelling the nonzero factor $\Lambda$ gives
\begin{equation}
    (c+d-1)G_{p,q}(c,d)
    =
    cG_{p+\frac12,q}(c-1,d)
    +
    dG_{p,q+\frac12}(c,d-1).
    \label{eq:G-transvection-recursion}
\end{equation}
Taking $(c,d)=(2,0)$ gives
\begin{equation}
    \beta_{p,q}=\beta_{p+\frac12,q}.
    \label{eq:beta-propagation}
\end{equation}

The \textbf{global seed} fixes the first-step coefficients at $p=\frac32$.
Indeed, comparison with \eqref{eq:H-action-W} gives
\begin{equation}
    \alpha_{\frac32,q}=q-1,
    \qquad
    \beta_{\frac32,q}=-\Lambda(q-2).
    \label{eq:first-step-seed}
\end{equation}
Since the full half-integer lattice is connected by the shifts in
\eqref{eq:alpha-propagation} and \eqref{eq:beta-propagation},
\begin{equation}
    \alpha_{p,q}=q-1,
    \qquad
    \beta_{p,q}=-\Lambda(q-2)
    \qquad
    \text{for all }p,q\in\tfrac12\mathbb Z.
\end{equation}
In particular, the exceptional first-channel term at $p=q=1$ and
second-channel term at $p=q=2$ vanish by these recursions.

Returning to the original indices,
\begin{equation}
    a(q-1)-b(p-1)
    =
    \bar m(q-1)-\bar n(p-1),
\end{equation}
and
\begin{equation}
    c(q-2)-d(p-2)
    =
    m(q-2)-n(p-2).
\end{equation}
The bracket is therefore uniquely fixed to
\begin{equation}
    \boxed{
    \begin{aligned}
    [W^p_{\bar m,m},W^q_{\bar n,n}]
    ={}&
    [\bar m(q-1)-\bar n(p-1)]\,
    W^{p+q-2}_{\bar m+\bar n,m+n}
    \\
    &-
    \Lambda[m(q-2)-n(p-2)]\,
    W^{p+q-1}_{\bar m+\bar n,m+n}.
    \end{aligned}
    }
    \label{eq:soft-algebra-derived}
\end{equation}
Substitution of the ten modes
\eqref{eq:global-identification} reproduces the complete geometric
action, including \eqref{eq:HH-global}.

\subsection{Jacobi identity for arbitrary modes}
\label{sec:full-Jacobi}

The preceding bootstrap determines the bracket for every pair
of wedge modes using Jacobi identities involving global
generators. We now verify that the resulting expression also
satisfies Jacobi for three arbitrary tower modes. This is a
consistency check of the derived bracket. Let
\begin{equation}
    \Phi(W^p_{\bar m,m})
    =
    x_1^{p-1+\bar m}x_2^{p-1-\bar m}
    y_1^{2-p+m}y_2^{2-p-m}.
    \label{eq:auxiliary-Laurent-map}
\end{equation}
These are homogeneous Laurent monomials of total degree two. Equip their Laurent algebra with the constant bracket
\begin{align}
    [f,g]_{\mathrm{aux}}
    ={}
    \frac12
    \left(
        \partial_{x_1}f\,\partial_{x_2}g
        -
        \partial_{x_2}f\,\partial_{x_1}g
    \right)+
    \frac{\Lambda}{2}
    \left(
        \partial_{y_1}f\,\partial_{y_2}g
        -
        \partial_{y_2}f\,\partial_{y_1}g
    \right).
    \label{eq:auxiliary-Poisson}
\end{align}
Direct differentiation gives
\begin{equation}
    [\Phi(W^p_{\bar m,m}),\Phi(W^q_{\bar n,n})]_{\mathrm{aux}}
    =
    \Phi\!\left([W^p_{\bar m,m},W^q_{\bar n,n}]\right),
\end{equation}
with the bracket on the right given by
\eqref{eq:soft-algebra-derived}.
For arbitrary modes $X,Y,Z$, compatibility of $\Phi$ with the bracket gives
\begin{equation}
    \Phi([X,[Y,Z]])
    =
    [\Phi(X),[\Phi(Y),\Phi(Z)]_{\rm aux}]_{\rm aux}.
\end{equation}
Hence
\begin{align}
    \Phi\!\left(
        [X,[Y,Z]]
        +[Y,[Z,X]]
        +[Z,[X,Y]]
    \right)
    =
    [\Phi(X),[\Phi(Y),\Phi(Z)]_{\rm aux}]_{\rm aux}
    +\text{cyclic}.
\end{align}
The right-hand side vanishes because
$[\cdot,\cdot]_{\rm aux}$ is a Poisson bracket. Therefore
\begin{equation}
    \Phi(\mathrm{Jacobiator})=0.
\end{equation}
 The map $\Phi$ is injective and a bracket homomorphism. 
Thus $\Phi(\mathrm{Jacobiator})=0$ implies
$\mathrm{Jacobiator}=0$, proving Jacobi for arbitrary Laurent modes.

The auxiliary realization establishes two separate facts. First,
the full Laurent-mode algebra satisfies the Jacobi identity. Second,
restricting to the wedge, for which
\begin{equation}
    a=p-1+\bar m\geq0,
    \qquad
    c=2-p+m\geq0,
\end{equation}
defines a closed subalgebra. A potentially negative output
index can occur only when $a=b=0$ in the first channel or $c=d=0$
in the second, and in either case the corresponding structure
coefficient vanishes. Hence we have Jacobi on the full Laurent algebra and closure of the wedge subalgebra.

    \section*{Acknowledgements}
     I thank Brian Kent for useful discussions. The work of H.K. is supported by CNS Spark Grant 2025-2029.

\appendix
\section{Spinors, real forms, and the AdS generator basis}
\label{app:spinors-real-forms}

The mode construction uses two commuting algebras, denoted
$\mathfrak{sl}(2)_L$ and $\mathfrak{sl}(2)_R$, together with four
AdS transvections. We summarize their spinor realization and explain
how the complex basis used in the calculation relates to Lorentzian
reality, the split real form, and the conventions of
~\cite{StromingerWei:2026}.

\subsection{Spinors and reality conditions}

Following Nagaraj et al.~\cite{NagarajPonomarev:2019,NagarajPonomarev:2020},
we parametrize null momenta by a pair of two-component spinors,
\begin{equation}
    k_{\alpha\dot\alpha}
    =
    \lambda_\alpha\widetilde\lambda_{\dot\alpha}.
\end{equation}
Here $\widetilde\lambda_{\dot\alpha}$ denotes the dotted spinor
(called $\bar\lambda_{\dot\alpha}$ in
refs.~\cite{NagarajPonomarev:2019,NagarajPonomarev:2020}).
For real Lorentzian null momentum one imposes the usual reality
condition
\begin{equation}
    \widetilde\lambda_{\dot\alpha}
    =
    (\lambda_\alpha)^*,
\end{equation}
up to the standard little-group rescaling. Then
$k_{\alpha\dot\alpha}$ is a Hermitian bispinor and represents a
real null vector in signature $(3,1)$.

For the algebraic analysis, however, it is convenient to complexify
the spinor variables and regard $\lambda_\alpha$ and
$\widetilde\lambda_{\dot\alpha}$ as independent. Then,
the complexified Lorentz algebra decomposes as
\begin{equation}
    \mathfrak{so}(3,1)_{\mathbb C}
    \simeq
    \mathfrak{sl}_2(\mathbb C)_L
    \oplus
    \mathfrak{sl}_2(\mathbb C)_R ,
    \label{eq:app-complex-Lorentz}
\end{equation}
with $\lambda_\alpha$ and $\widetilde\lambda_{\dot\alpha}$
transforming under the two factors, respectively. The dotted and
undotted sectors are related by complex conjugation only after the
Lorentzian reality condition is imposed.

This complexification should not be confused with a change to split
signature. If instead both spinors are taken to be independently
real, the bispinor $k_{\alpha\dot\alpha}$ is a real $2\times2$
matrix, whose determinant defines a quadratic form of signature
$(2,2)$. The corresponding real spin algebra is then
\begin{equation}
    \mathfrak{so}(2,2)
    \simeq
    \mathfrak{sl}_2(\mathbb R)_L
    \oplus
    \mathfrak{sl}_2(\mathbb R)_R .
\end{equation}
Thus, complexifying the Lorentzian spinor variables and choosing the
split-signature real form are different operations. In what follows
we use the complexified spinor representation as an analytic
convenience, while the physical AdS$_4$ geometry and its Dirichlet
boundary condition remain Lorentzian.

\subsection{Global generators and their action on modes}

Write $\lambda=(u,v)$ and
$\widetilde\lambda=(\widetilde u,\widetilde v)$. A convenient
normalization of the first factor is
\begin{equation}
 L_{-1}=-u\partial_v,\qquad
 L_0=\tfrac12(u\partial_u-v\partial_v),\qquad
 L_1=v\partial_u,
 \label{eq:app-six-L-generators}
\end{equation}
with $\bar L_r$ given by the same expressions in dotted variables.
The four remaining generators are the AdS transvections,
\begin{equation}
 \mathcal P_{\alpha\dot\alpha}
 =\lambda_\alpha\widetilde\lambda_{\dot\alpha}
 +\Lambda\frac{\partial}{\partial\lambda^\alpha}
          \frac{\partial}{\partial\widetilde\lambda^{\dot\alpha}},
 \qquad
 H_{\bar r_{\dot\alpha},r_\alpha}
 =-\tfrac12\mathcal P_{\alpha\dot\alpha},
 \label{eq:app-NP-generators}
\end{equation}
where $\Lambda=-\ell^{-2}$ is our curvature parameter. The transvection generator follows this spinor index convention
\begin{equation}
    \alpha=1\ \longleftrightarrow\ r=-\frac12,
    \qquad
    \alpha=2\ \longleftrightarrow\ r=+\frac12,
\end{equation}
and analogously
$\dot\alpha=\dot1,\dot2\leftrightarrow
\bar r=-\tfrac12,+\tfrac12$.
Thus the mode labels $r,\bar r$ are simply a relabelling of the
undotted and dotted spinor indices. 
 We raise spinor indices with $\epsilon^{12}=1$,
$\epsilon_{12}=-1$. Thus
$\partial/\partial\lambda^1=\partial_v$ and
$\partial/\partial\lambda^2=-\partial_u$.
These operators obey
\begin{align}
 [L_k,L_l]&=(k-l)L_{k+l},&
 [\bar L_k,\bar L_l]&=(k-l)\bar L_{k+l},&
 [L_k,\bar L_l]&=0,
 \nonumber\\
 [L_k,H_{\bar r,r}]
 &=(\tfrac{k}{2}-r)H_{\bar r,r+k},&
 [\bar L_k,H_{\bar r,r}]
 &=(\tfrac{k}{2}-\bar r)H_{\bar r+k,r},
 \label{eq:app-two-sl2}\\
 [H_{\bar r,r},H_{\bar s,s}]
 &=\frac{\Lambda}{2}\left[
 (\bar r-\bar s)L_{r+s}
 +(r-s)\bar L_{\bar r+\bar s}\right].
 \label{eq:app-full-global-algebra}
\end{align}
Here $k,l=0,\pm1$ and
$r,\bar r,s,\bar s=\pm\tfrac12$. Whenever a shifted label lies outside these ranges, the corresponding prefactor vanishes.
The transvections transform as $(\mathbf2,\mathbf2)$ under the two
factors and close the ten-generator complex algebra
$\mathfrak{so}(5,\mathbb C)$. Their commutator is proportional to
curvature, as expected for the AdS counterparts of translations.

To pass to projective variables, set
$\lambda=a(1,z)$,
$\widetilde\lambda=b(1,\bar z)$ and $\omega=ab$.
On helicity-$\sigma$ fields, the operator
$\tfrac12(b\partial_b-a\partial_a)$ has eigenvalue $\sigma$.
A Mellin transform in $\omega$ replaces
$\omega\partial_\omega$ by $-\Delta$ giving
\begin{equation}
 \begin{aligned}
 \mathscr D_r^{(h)}
 &=-z^{r+1}\partial_z-(r+1)h z^r,
 &h&=\tfrac12(\Delta+\sigma),\\
 \bar{\mathscr D}_r^{(\bar h)}
 &=-\bar z^{r+1}\partial_{\bar z}
   -(r+1)\bar h\bar z^r,
 &\bar h&=\tfrac12(\Delta-\sigma).
 \end{aligned}
 \label{eq:app-projective-after-Mellin}
\end{equation}
These weights describe spinor homogeneity, not the dimension of a
three-dimensional boundary primary. The transvections preserve
helicity but mix Mellin sectors through their multiplication and
derivative terms.

For $\sigma=2$ and $\Delta=4-2p$, the field weights are
$(\bar h,h)=(1-p,3-p)$. When $2p-2$ is a nonnegative integer,
the barred polynomials of degree at most $2p-2$ form a finite
submodule.


\subsection{Real subgroups and the Strominger-Wei basis}

It is useful to distinguish two different issues here. The choice of a
real form of the six-generator
$\mathfrak{sl}_2(\mathbb C)_L\oplus\mathfrak{sl}_2(\mathbb C)_R$
subalgebra, and the embedding of the corresponding real subgroup
inside the physical Lorentzian AdS$_4$ isometry group
$SO(3,2)$. The full complex algebra discussed above is
$\mathfrak{so}(5,\mathbb C)$, whose split real form is
$\mathfrak{so}(3,2)$.

Consider the ambient realization with metric
\begin{equation}
    \eta=\operatorname{diag}(-1,1,1,1,-1).
\end{equation}
The Lorentz subgroup $SO(3,1)\subset SO(3,2)$ may be realized
as the stabilizer of the timelike direction $e_4$, whereas an
$SO(2,2)$ subgroup is obtained as the stabilizer of the spacelike
direction $e_3$. In terms of ambient generators
$\mathsf J_{AB}=X_A\partial_B-X_B\partial_A$,
\begin{equation}
\begin{aligned}
    \mathfrak h_{\rm Lor}
    &=
    \operatorname{span}_{\mathbb R}
    \{\mathsf J_{ab}:a,b=0,1,2,3\}
    \simeq\mathfrak{so}(3,1),\\
    \mathfrak h_{\rm split}
    &=
    \operatorname{span}_{\mathbb R}
    \{\mathsf J_{01},\mathsf J_{02},\mathsf J_{12},
      \mathsf J_{04},\mathsf J_{14},\mathsf J_{24}\}
    \simeq\mathfrak{so}(2,2).
\end{aligned}
\label{eq:app-geometric-so22}
\end{equation}
Relative to the first decomposition, the second subgroup contains
three Lorentz generators and three AdS transvections. The two real
subgroups are not conjugate inside $SO(3,2)$, since a real AdS
isometry cannot exchange timelike and spacelike ambient directions.
Their complexifications are conjugate inside $SO(5,\mathbb C)$.
 Taking the dotted and undotted spinors
independently real selects the split real form
\begin{equation}
    \mathfrak{sl}_2(\mathbb R)_L
    \oplus
    \mathfrak{sl}_2(\mathbb R)_R
    \simeq
    \mathfrak{so}(2,2)
\end{equation}
of the complexified six-generator subalgebra. But it does not
specify its geometric embedding inside the fixed Lorentzian
$SO(3,2)$ real form.

For comparison, denote the conformal Killing fields called
$L_r$, $\bar L_r$ and $H_{\bar r,r}$ in
ref.~\cite{StromingerWei:2026} (Sec.~3), by
$\zeta_r$, $\bar\zeta_r$ and $\xi_{\bar r,r}$.
Our normalization is
\begin{equation}
    L_r=i\zeta_r,
    \qquad
    \bar L_r=i\bar\zeta_r,
    \qquad
    H_{\bar r,r}
    =
    \frac{i}{\sqrt2\,\ell}\xi_{\bar r,r},
    \label{eq:app-SW-normalization}
\end{equation}
with $\Lambda=-\ell^{-2}$.

Their explicit Lorentzian vector fields obey
\begin{equation}
    \zeta_r^*=-\bar\zeta_r,
    \qquad
    \xi_{\bar r,r}^*=\xi_{r,\bar r},
\end{equation}
and therefore, in our normalization,
\begin{equation}
    L_r^*=\bar L_r,
    \qquad
    H_{\bar r,r}^*=-H_{r,\bar r}.
    \label{eq:app-Lorentz-reality}
\end{equation}
The six-generator real algebra selected by this conjugation is
$\mathfrak{so}(3,1)$.

Imposing the separate reality conditions instead
\begin{equation}
    L_r^*=L_r,
    \qquad
    \bar L_r^*=\bar L_r,
\end{equation}
selects
\begin{equation}
    \mathfrak{sl}_2(\mathbb R)_L
    \oplus
    \mathfrak{sl}_2(\mathbb R)_R
    \simeq
    \mathfrak{so}(2,2),
\end{equation}
and hence a different real structure.

The comparison of modes and commutators is therefore most naturally
made at the level of the complexified algebra. The Lorentzian reality
condition on the physical graviton is imposed separately. In
particular, the Lorentzian Hodge operator acting on two-forms obeys
\begin{equation}
    \star^2=-1.
\end{equation}
\section{AdS transvection action on graviton modes}
\label{app:adstranvection}

We derive the transvection action directly from the spinor-helicity
generator
\begin{equation}
    \mathcal P_{\alpha\dot\alpha}
    =
    \lambda_\alpha\bar\lambda_{\dot\alpha}
    +
    \Lambda
    \frac{\partial}{\partial\lambda^\alpha}
    \frac{\partial}{\partial\bar\lambda^{\dot\alpha}},
    \qquad
    H_{\bar r,r}=-\frac12\mathcal P_{\alpha\dot\alpha}.
    \label{eq:P-direct}
\end{equation}
Write
\begin{equation}
    \lambda_\alpha=a(1,z),
    \qquad
    \bar\lambda_{\dot\alpha}=b(1,\bar z),
    \qquad
    \omega=ab,
    \qquad
    \rho=\frac{a}{b}.
\end{equation}
We identify
\begin{equation}
    \alpha=1,2
    \quad\longleftrightarrow\quad
    r=-\frac12,+\frac12,
    \qquad
    \dot\alpha=\dot1,\dot2
    \quad\longleftrightarrow\quad
    \bar r=-\frac12,+\frac12,
\end{equation}
and define
\begin{equation}
    q_r(z)=z^{\,r+\frac12},
    \qquad
    \bar q_{\bar r}(\bar z)
    =\bar z^{\,\bar r+\frac12}.
\end{equation}
The multiplication part of \eqref{eq:P-direct} is therefore
\begin{equation}
    \lambda_\alpha\bar\lambda_{\dot\alpha}
    =
    \omega\,q_r(z)\bar q_{\bar r}(\bar z).
    \label{eq:P-multiplication}
\end{equation}

For the derivative part, using
$\epsilon^{12}=1$ gives
\begin{align}
    \frac{\partial}{\partial\lambda^1}
    &=
    \frac1a\partial_z,
    &
    \frac{\partial}{\partial\lambda^2}
    &=
    \frac1a
    \left(
        -\omega\partial_\omega
        -\rho\partial_\rho
        +z\partial_z
    \right),
    \\
    \frac{\partial}{\partial\bar\lambda^{\dot1}}
    &=
    \frac1b\partial_{\bar z},
    &
    \frac{\partial}{\partial\bar\lambda^{\dot2}}
    &=
    \frac1b
    \left(
        -\omega\partial_\omega
        +\rho\partial_\rho
        +\bar z\partial_{\bar z}
    \right).
    \label{eq:spinor-derivatives-projective}
\end{align}
For a helicity-$\sigma$ field,
$\rho\partial_\rho\Psi_\sigma=-\sigma\Psi_\sigma$.
Keeping the $\rho$ dependence until both derivatives have acted, the
Mellin transform gives
\begin{equation}
    \int_0^\infty d\omega\,
    \omega^{\Delta-1}
    \frac{\partial^2\Psi_\sigma}
         {\partial\lambda^\alpha
          \partial\bar\lambda^{\dot\alpha}}
    =
    \mathscr D_r\,
    \bar{\mathscr D}_{\bar r}\,
    \widetilde\Psi_{\Delta-1,\sigma},
    \label{eq:derivative-mellin-action}
\end{equation}
where
\begin{align}
    \mathscr D_r
    &=
    z^{\,r+\frac12}\partial_z
    +
    \left(r+\frac12\right)
    (\Delta-1+\sigma)\,
    z^{\,r-\frac12},
    \\
    \bar{\mathscr D}_{\bar r}
    &=
    \bar z^{\,\bar r+\frac12}\partial_{\bar z}
    +
    \left(\bar r+\frac12\right)
    (\Delta-1-\sigma)\,
    \bar z^{\,\bar r-\frac12}.
    \label{eq:D-projective}
\end{align}
Here we used
\begin{equation}
    \int_0^\infty d\omega\,
    \omega^{\Delta-2}
    \omega\partial_\omega\Psi
    =
    -(\Delta-1)
    \int_0^\infty d\omega\,
    \omega^{\Delta-2}\Psi .
\end{equation}

For the positive-helicity graviton,
$\sigma=2$ and $\Delta=4-2p$. Hence
\begin{align}
    [H_{\bar r,r},\mathcal H^{p,+}]
    ={}&
    -\frac12\,
    q_r\bar q_{\bar r}\,
    \mathcal H^{p-\frac12,+}
    \nonumber\\
    &-
    \frac{\Lambda}{2}\,
    \mathscr D_r\bar{\mathscr D}_{\bar r}\,
    \mathcal H^{p+\frac12,+},
    \label{eq:H-generating-field-action}
\end{align}
with
\begin{align}
    \mathscr D_r
    &=
    z^{\,r+\frac12}\partial_z
    +
    \left(r+\frac12\right)(5-2p)
    z^{\,r-\frac12},
    \\
    \bar{\mathscr D}_{\bar r}
    &=
    \bar z^{\,\bar r+\frac12}\partial_{\bar z}
    +
    \left(\bar r+\frac12\right)(1-2p)
    \bar z^{\,\bar r-\frac12}.
    \label{eq:D-p}
\end{align}

We now extract the Laurent coefficients using
\begin{equation}
    w^p_{\bar m,m}
    =
    N_{p,\bar m}
    \oint\frac{dz}{2\pi i}\,
    z^{m+2-p}
    \oint\frac{d\bar z}{2\pi i}\,
    \bar z^{\bar m-p}
    \mathcal H^{p,+},
\end{equation}
with
\begin{equation}
    N_{p,\bar m}
    =
    (-1)^{p-1-\bar m}
    \Gamma(p+\bar m)\Gamma(p-\bar m).
    \label{eq:N-transvection}
\end{equation}

For the multiplication term, the contour selects
\begin{equation}
    (\bar n,n)
    =
    (\bar m+\bar r,m+r)
\end{equation}
at level $p-\frac12$. The normalization ratio is
\begin{equation}
    \frac{N_{p,\bar m}}
         {N_{p-\frac12,\bar m+\bar r}}
    =
    \bar m-2\bar r(p-1).
\end{equation}
Therefore
\begin{equation}
    -\frac12
    \frac{N_{p,\bar m}}
         {N_{p-\frac12,\bar m+\bar r}}
    =
    \bar r(p-1)-\frac{\bar m}{2}
    \equiv A_p(\bar r,\bar m).
    \label{eq:A-direct}
\end{equation}

For the derivative term, consider the selected monomial of
$\mathcal H^{p+\frac12,+}$,
\begin{equation}
    \bar z^{\,p-\frac12-\bar m-\bar r}\,
    z^{\,p-\frac52-m-r}.
\end{equation}
The unbarred operator gives
\begin{equation}
    \mathscr D_r
    z^{\,p-\frac52-m-r}
    =
    -2
    \left[
        r(p-2)+\frac{m}{2}
    \right]
    z^{\,p-3-m},
    \label{eq:D-unbar-action}
\end{equation}
while
\begin{equation}
    \bar{\mathscr D}_{\bar r}
    \bar z^{\,p-\frac12-\bar m-\bar r}
    =
    -2\bar r
    \left(
        p+2\bar r\bar m
    \right)
    \bar z^{\,p-1-\bar m}.
    \label{eq:D-bar-action}
\end{equation}
The corresponding normalization ratio is
\begin{equation}
    \frac{N_{p,\bar m}}
         {N_{p+\frac12,\bar m+\bar r}}
    =
    \frac{1}
    {2\bar r\left(p+2\bar r\bar m\right)}.
    \label{eq:N-derivative-ratio}
\end{equation}
The barred factor therefore cancels against the normalization ratio,
leaving
\begin{equation}
    \frac{N_{p,\bar m}}
         {N_{p+\frac12,\bar m+\bar r}}\,
    \mathscr D_r\bar{\mathscr D}_{\bar r}
    \longrightarrow
    2
    \left[
        r(p-2)+\frac{m}{2}
    \right].
\end{equation}
Multiplication by the prefactor $-\Lambda/2$ in
\eqref{eq:H-generating-field-action} gives
\begin{equation}
    -\Lambda
    \left[
        r(p-2)+\frac{m}{2}
    \right]
    \equiv
    -\Lambda B_p(r,m).
    \label{eq:B-direct}
\end{equation}

Thus the transvection action is
\begin{equation}
    \boxed{
    \begin{aligned}
    [H_{\bar r,r},w^p_{\bar m,m}]
    ={}&
    \left[
        \bar r(p-1)-\frac{\bar m}{2}
    \right]
    w^{p-\frac12}_{\bar m+\bar r,m+r}
    \\
    &-
    \Lambda
    \left[
        r(p-2)+\frac{m}{2}
    \right]
    w^{p+\frac12}_{\bar m+\bar r,m+r}.
    \end{aligned}
    }
    \label{eq:H-action-direct}
\end{equation}
Hence
\begin{equation}
    A_p(\bar r,\bar m)
    =
    \bar r(p-1)-\frac{\bar m}{2},
    \qquad
    B_p(r,m)
    =
    r(p-2)+\frac{m}{2}.
\end{equation}
\section{Metric potentials and their Weyl tensors}
\label{subsec:ads_metric_potentials}

The AdS$_4$ metric can be written as a Weyl factor times the flat space metric.
\begin{equation}
    ds^2
    =
    \Omega^2(x)\eta_{\mu\nu}dx^\mu dx^\nu,
    \qquad
    \Omega(x)=G(x)^{-1}, \quad  G(x)=1-\frac{x^2}{4\ell^2}
    \label{eq:ads_stereographic_metric}
\end{equation}
Using these spinors, one can define a null bispinor as
\begin{equation}
    k_{\alpha\dot\alpha}
    =
    \lambda_\alpha\bar\lambda_{\dot\alpha}.
    \label{eq:null_k}
\end{equation}
Using the spinors and momenta, we can write the regular spin-two field strengths as
\begin{align}
    C^{(+)}_{\dot\alpha\dot\beta\dot\gamma\dot\delta}
    (x;\lambda,\bar\lambda)
    &=-\frac{1}{2}
    \bar\lambda_{\dot\alpha}
    \bar\lambda_{\dot\beta}
    \bar\lambda_{\dot\gamma}
    \bar\lambda_{\dot\delta}\,
    G^3(x)e^{ik\cdot x},
    \label{eq:positive_ads_weyl}\\
    C^{(-)}_{\alpha\beta\gamma\delta}
    (x;\lambda,\bar\lambda)
    &=
    -\frac{1}{2}\lambda_\alpha\lambda_\beta
    \lambda_\gamma\lambda_\delta\,
    G^3(x)e^{ik\cdot x}.
    \label{eq:negative_ads_weyl}
\end{align}
The positive-helicity field satisfies
\begin{equation}
    \nabla^{\alpha\dot\alpha}
    C^{(+)}_{\dot\alpha\dot\beta\dot\gamma\dot\delta}=0,
    \label{eq:weyl_eom_plus}
\end{equation}
with the corresponding equation in the opposite-helicity sector.
Their little-group scaling agrees with
Eq.~\eqref{eq:little_group_ads}.  One can even write the metric
$h^{(\sigma)}_{\mu\nu}(x;\lambda,\bar\lambda)$ whose nonzero
linearized Weyl components are precisely
Eqs.~\eqref{eq:positive_ads_weyl} and
\eqref{eq:negative_ads_weyl}.
Expanding
$g_{\mu\nu}=g^{\mathrm{AdS}}_{\mu\nu}
+\varepsilon h^{(\sigma)}_{\mu\nu}$,
These fluctuations obey the linearized Einstein eq.
\begin{equation}
    \mathcal E_{\mu\nu}[h^{(\sigma)}]
    \equiv
    \delta G_{\mu\nu}[h^{(\sigma)}]
    +
    \Lambda_{\rm Ein}h^{(\sigma)}_{\mu\nu}
    =0,
    \qquad
    \Lambda_{\rm Ein}=-\frac{3}{\ell^2}.
    \label{eq:linearized-einstein}
\end{equation}
The metric representatives are
defined modulo
\begin{equation}
    h_{\mu\nu}
    \longrightarrow
    h_{\mu\nu}+2\nabla_{(\mu}\xi_{\nu)}.
    \label{eq:linearized_diff}
\end{equation}
The linearized Weyl tensor is invariant under this transformation
because the AdS background Weyl tensor vanishes.

\paragraph{Metric potential:-}
Let $\mu_\alpha$ be a reference spinor and introduce
\begin{equation}
    \mathfrak a
    =\lambda_\alpha\bar\lambda_{\dot\alpha}
      x^{\dot\alpha\alpha}
    =-2k\cdot x,
    \qquad
    \mathfrak b
    =x_{\alpha\dot\alpha}x^{\dot\alpha\alpha}
    =-2x^2,
    \qquad
    G=1+\frac{\mathfrak b}{8\ell^2}.
\end{equation}
The positive-helicity metric potential of
\cite{NagarajPonomarev:2020}, (Eq.~(6.22)) is
\begin{align}
    h^{(+)}_{\alpha\dot\alpha,\beta\dot\beta}
    ={}&
    -\left(
        G-\frac{i\mathfrak b}{2\ell^2\mathfrak a}
    \right)
    \frac{
        \mu_\alpha\mu_\beta
        \bar\lambda_{\dot\alpha}\bar\lambda_{\dot\beta}
    }{
        \langle\mu\lambda\rangle^2
    }e^{ik\cdot x}
    \nonumber\\
    &-
    \frac{i e^{ik\cdot x}}{2\ell^2\mathfrak a}
    \frac{
        \mu_\alpha\mu_\beta
        \left(
            \lambda^\gamma x_{\gamma\dot\alpha}
            \bar\lambda_{\dot\beta}
            +
            \lambda^\gamma x_{\gamma\dot\beta}
            \bar\lambda_{\dot\alpha}
        \right)
    }{
        \langle\mu\lambda\rangle^3
    }
    \langle\mu x\lambda].
    \label{eq:NP_metric_potential}
\end{align}
These are local frame components. Spacetime components
$h_{\mu\nu}$ follow by inserting the background vierbeins.

Using the curvature operator in Eq.~(6.17) of that reference gives (with the on-shell conversion $C^{(1)}=-\tfrac12 F$, gives)
\begin{align}
    C^{(1)}_{\dot\alpha\dot\beta\dot\gamma\dot\delta}
    [h^{(+)}]
    &=\frac{-1}{2}
    \bar\lambda_{\dot\alpha}\bar\lambda_{\dot\beta}
    \bar\lambda_{\dot\gamma}\bar\lambda_{\dot\delta}
    G^3 e^{ik\cdot x},
    \label{eq:potential_reproduces_weyl}\\
    C^{(1)}_{\alpha\beta\gamma\delta}[h^{(+)}]
    &=0.
\end{align}
The Ricci components of the gauge-invariant curvature also vanish. And it satisfies the Einstein eq.
which gives
\begin{equation}
    \mathcal E_{\mu\nu}[h^{(+)}]
    \equiv
    \delta G_{\mu\nu}[h^{(+)}]
    +\Lambda_{\rm Ein}h^{(+)}_{\mu\nu}
    =0,
    \qquad
    \Lambda_{\rm Ein}=-\frac{3}{\ell^2}.
    \label{eq:linearized-einstein}
\end{equation}
The opposite helicity follows by exchanging dotted and undotted
spinors.

\subsection{Mellin inversion and discrete homogeneous sectors}
\label{subsec:mellin_contour_soft_weights}
We start with spinors as
\begin{equation}
    \lambda_\alpha=\sqrt{\omega}\,q_\alpha(z),
    \qquad
    \bar\lambda_{\dot\alpha}
        =\sqrt{\omega}\,\bar q_{\dot\alpha}(\bar z),
    \qquad
    k_{\alpha\dot\alpha}
        =\omega Q_{\alpha\dot\alpha},
    \qquad
    Q_{\alpha\dot\alpha}=q_\alpha\bar q_{\dot\alpha}.
\end{equation}
Keeping the reference spinor fixed, we can count the scale in
Eq.~\eqref{eq:NP_metric_potential} and organize it in terms of scale $\omega$ as
\begin{equation}
    h^{(\sigma)}_{\mu\nu}(x;\omega,z,\bar z)
    =
    e^{i\omega Q\cdot x}
    \left(
        H^{(0,\sigma)}_{\mu\nu}
        +\frac{H^{(-1,\sigma)}_{\mu\nu}}{\omega}
    \right),
    \label{eq:potential_scale_dependence}
\end{equation}
where both coefficient tensors are independent of $\omega$.
The second term is curvature dependent.

Introduce a regulator $\epsilon>0$ and define the Mellin transform as
\begin{equation}
    \Psi^{\Delta,\sigma}_{\mu\nu;\epsilon}
    =
    \int_0^\infty d\omega\,
    \omega^{\Delta-1}e^{-\epsilon\omega}
    h^{(\sigma)}_{\mu\nu}(x;\omega,z,\bar z).
    \label{eq:metric_mellin_transform}
\end{equation}
Writing $Z_\epsilon=\epsilon-iQ\cdot x$, the integral evaluates to
\begin{equation}
    \Psi^{\Delta,\sigma}_{\mu\nu;\epsilon}
    =
    \frac{\Gamma(\Delta)}{Z_\epsilon^\Delta}
        H^{(0,\sigma)}_{\mu\nu}
    +
    \frac{\Gamma(\Delta-1)}{Z_\epsilon^{\Delta-1}}
        H^{(-1,\sigma)}_{\mu\nu}.
    \label{eq:metric_mellin_explicit}
\end{equation}
For this representative, $\operatorname{Re}\Delta>1$ is a
half-plane of absolute convergence. The regulator controls
$\omega\to\infty$, while the $\omega^{-1}$ term controls
$\omega\to0$.  Equation~\eqref{eq:metric_mellin_explicit}
then supplies the meromorphic continuation in $\Delta$.

Since the Einstein and Weyl operators act only on spacetime,
\begin{equation}
    \mathcal E[\Psi^{\Delta,\sigma}_{\epsilon}]=0,
    \qquad
    C^{(1)}[\Psi^{\Delta,\sigma}_{\epsilon}]
    =
    \int_0^\infty d\omega\,
    \omega^{\Delta-1}e^{-\epsilon\omega}
    C^{(1)}[h^{(\sigma)}(\omega)].
    \label{eq:mellin_commutes_with_eom}
\end{equation}
The metric and curvature constructions therefore give the same
Mellin-transformed on-shell field.

The inverse transform is
\begin{equation}
    e^{-\epsilon\omega}h^{(\sigma)}_{\mu\nu}(\omega)
    =
    \frac{1}{2\pi i}
    \int_{c-i\infty}^{c+i\infty}
    d\Delta\,
    \omega^{-\Delta}
    \Psi^{\Delta,\sigma}_{\mu\nu;\epsilon},
    \qquad c>1.
    \label{eq:metric_inverse_mellin}
\end{equation}
Indeed, with $\omega=e^u$ and $\Delta=c+i\nu$, the Mellin
transform is the Fourier transform of
$e^{cu-\epsilon e^u}h^{(\sigma)}_{\mu\nu}(e^u)$.
Fourier inversion proves
eq.~\eqref{eq:metric_inverse_mellin}.
Thus invertibility requires a suitable vertical contour, not a
discrete set of Mellin weights.  In the flat limit the
$\omega^{-1}$ term disappears and the convergence condition
relaxes to $c>0$. Pasterski and Shao~\cite{PasterskiShao:2017} have shown the principal series representation for the flat space gravitons as $\Delta=1+i\nu,\quad \nu\in\mathbb R$.

\paragraph{The discrete soft sectors.}

The soft algebra uses a different prescription. The analytic continuation
of the homogeneous family followed by the discrete selection
\begin{equation}
    \Delta_p=4-2p,
    \qquad
    p\in\tfrac12\mathbb Z.
    \label{eq:soft_mellin_lattice}
\end{equation}
For $\sigma=+2$, it gives
\begin{equation}
    (\bar h_{\rm gen},h_{\rm gen})=(1-p,3-p).
    \label{eq:generating_weights_p}
\end{equation}
For $p=1,\frac32,2,\ldots$, the barred representation contains the
finite polynomial submodule of degree $2p-2$.
Extending the same labeling to all half-integer $p$ defines the
discrete lattice used in the full Laurent completion. The same distinction underlies the integer
conformally soft weights in flat space
\cite{GuevaraEtAl:2021,Strominger:2021}.

\subsection{Metric realization of the Dirichlet pairing}
\label{sec:metric-dirichlet-pairing}

We now verify directly that the helicity combination
\begin{equation}
    W^p_{\bar m,m}
    =
    w^p_{\bar m,m}
    +
    \Lambda^{3-2p}\,
    \bar w^{\,3-p}_{\bar m,m}
    \label{eq:dirichlet-pair-recall}
\end{equation}
does satisfy the Dirichlet boundary condition.  The calculation being done here is in a
flat conformal boundary patch.  We first work in the
transverse-traceless sector in which the linearized Cotton operator is
invertible. 

In Fefferman-Graham gauge, write
\begin{equation}
    ds^2
    =
    \frac{\ell^2}{r^2}
    \left[
        dr^2+
        \bigl(\eta_{ij}+\gamma_{ij}\bigr)dy^i dy^j
    \right],
    \qquad
    H_{rr}=H_{ri}=0 .
\end{equation}
For a transverse-traceless linearized Einstein perturbation,
\begin{equation}
    \gamma_{ij}''
    -\frac{2}{r}\gamma_{ij}'
    +\Box\gamma_{ij}
    =0,
\end{equation}
and near the boundary
\begin{equation}
    \gamma_{ij}
    =
    s_{ij}
    +\frac{r^2}{2}\Box s_{ij}
    +r^3t_{ij}
    +O(r^4).
    \label{eq:FG-linear-expansion}
\end{equation}
Thus
\begin{equation}
    \delta g_{(0)ij}=s_{ij}.
\end{equation}
Let $E_{ij}$ and $B_{ij}$ denote the electric and magnetic parts of
the Weyl tensor of the conformally rescaled metric.  With the
conventions
\begin{equation}
    E_{ij}
    =
    -\frac14\gamma_{ij}''
    +\frac14\Box\gamma_{ij},
    \qquad
    B_{ij}
    =
    \frac12\epsilon_i{}^{kl}
    \partial_k\gamma_{lj}',
\end{equation}
their finite boundary limits are
\begin{equation}
    \mathcal E_{ij}
    \equiv
    \lim_{r\rightarrow0}r^{-1}E_{ij}
    =
    -\frac32 t_{ij},
    \qquad
    \mathcal B_{ij}
    \equiv
    \lim_{r\rightarrow0}r^{-1}B_{ij}
    =
    -\mathscr C[s]_{ij},
    \label{eq:EB-FG}
\end{equation}
where
\begin{equation}
    \mathscr C[s]_{ij}
    =
    -\frac12\epsilon_i{}^{kl}
    \partial_k\Box s_{lj}
\end{equation}
is the linearized Cotton operator.  These are the standard
electric and magnetic-Cotton relations
\cite{deHaro:2009}.

For completeness, Einstein's equation also gives
\begin{equation}
    \left(\frac{\gamma_{ij}'}{r}\right)'
    =
    -\frac{2E_{ij}}{r},
\end{equation}
and hence
\begin{equation}
    \gamma_{ij}(r)
    =
    s_{ij}
    +\frac{r^2}{2}\Box s_{ij}
    -
    \int_0^r du\,
    \frac{r^2-u^2}{u}\,
    E_{ij}(u).
    \label{eq:metric-from-electric}
\end{equation}
On the sector where $\mathscr C$ is invertible,
Eq.~\eqref{eq:EB-FG} therefore fixes the boundary source as
\begin{equation}
    s_{ij}
    =
    -\mathscr C^{-1}\mathcal B_{ij}.
    \label{eq:source-from-B}
\end{equation}

For the two chiral Weyl tensors,
\begin{equation}
    \mathcal B^{(+)}_{ij}
    =
    i\mathcal E^{(+)}_{ij},
    \qquad
    \mathcal B^{(-)}_{ij}
    =
    -i\mathcal E^{(-)}_{ij}.
\end{equation}
The boundary source carried by
Eq.~\eqref{eq:dirichlet-pair-recall} is
\begin{equation}
    \delta g_{(0)ij}
    [W^p_{\bar m,m}]
    =
    -i\mathscr C^{-1}
    \left[
        \mathcal E_+
        \bigl(w^p_{\bar m,m}\bigr)
        -
        \Lambda^{3-2p}
        \mathcal E_-
        \bigl(\bar w^{\,3-p}_{\bar m,m}\bigr)
    \right]_{ij}.
    \label{eq:source-electric-difference}
\end{equation}
It remains to verify that the two electric profiles appearing here
indeed agree with the required relative normalization.

\paragraph{Electric-profile matching.}
We can use the eq. \eqref{eq:positive_ads_weyl} and then take the Mellin transform of the Weyl tensor to find the electric components. \\

We evaluate the boundary electric data in a local conformal frame at the
boundary point
\begin{equation}
    x_\ast^\mu=(0,0,0,2\ell),
    \qquad
    G(x_\ast)=0,
\end{equation}
for which
\begin{equation}
    Q(z,\bar z)\cdot x_\ast
    =
    \ell(1-z\bar z).
\end{equation}
We use the projective spinors
\begin{equation}
    q_\alpha(z)=(1,z),
    \qquad
    \bar q_{\dot\alpha}(\bar z)=(1,\bar z),
\end{equation}
so that
\begin{equation}
    \lambda_\alpha=\sqrt{\omega}\,q_\alpha,
    \qquad
    \bar\lambda_{\dot\alpha}
    =\sqrt{\omega}\,\bar q_{\dot\alpha},
    \qquad
    k_{\alpha\dot\alpha}
    =
    \omega Q_{\alpha\dot\alpha},
    \qquad
    Q_{\alpha\dot\alpha}
    =
    q_\alpha\bar q_{\dot\alpha}.
\end{equation}
Let $n_{\alpha}{}^{\dot\alpha}$ denote the outward boundary normal,
viewed as the local bispinor with dotted and undotted spin index.  We choose the boundary spin frame at $x_\ast$ such that
\begin{equation}
    \widehat{\bar q}_\alpha
    \equiv
    n_{\alpha}{}^{\dot\alpha}\bar q_{\dot\alpha}
    =
    (\bar z,1)_\alpha ,
    \qquad
    q_\alpha=(1,z)_\alpha .
    \label{eq:boundary-spin-frame}
\end{equation}
The positive-helicity Weyl spinor is converted to a boundary symmetric
spinor by contracting each dotted index with the normal,
\begin{equation}
    \mathcal E^{(+)}_{\alpha\beta\gamma\delta}
    \propto
    n_{\alpha}{}^{\dot\alpha}
    n_{\beta}{}^{\dot\beta}
    n_{\gamma}{}^{\dot\gamma}
    n_{\delta}{}^{\dot\delta}
    \widetilde C^{(+)}_{\dot\alpha\dot\beta\dot\gamma\dot\delta},
    \label{eq:electric-spinor-projection-plus}
\end{equation}
whereas the negative-helicity Weyl spinor carries undotted
indices,
\begin{equation}
    \mathcal E^{(-)}_{\alpha\beta\gamma\delta}
    \propto
    \widetilde C^{(-)}_{\alpha\beta\gamma\delta}.
    \label{eq:electric-spinor-projection-minus}
\end{equation}
Thus, suppressing the common conversion factor between the Weyl spinor
and the finite Fefferman-Graham electric data 
\footnote{
The Weyl spinors of \cite{NagarajPonomarev:2020} are local-Lorentz
components.  With
$e_\mu{}^a=G^{-1}\delta_\mu^a$, their regular spin-two solution scales as
\[
    C^{\rm Lor}_{abcd}\sim G^3 .
\]
Hence, the corresponding Weyl tensor with four lower coordinate
indices scales as
\[
    C_{\mu\nu\rho\sigma}
    =
    e_\mu{}^a e_\nu{}^b e_\rho{}^c e_\sigma{}^d
    C^{\rm Lor}_{abcd}
    \sim G^{-1}.
\]
For the regular conformal metric
$\widehat g_{\mu\nu}=G^2g_{\mu\nu}$, conformal covariance of the Weyl
tensor gives
\[
    \widehat C_{\mu\nu\rho\sigma}
    =
    G^2 C_{\mu\nu\rho\sigma}
    \sim G .
\]
Hence its electric projection with respect to the finite conformal
normal behaves as
\[
    \widehat E_{ij}\sim G .
\]
Near the boundary point used below, the stereographic defining
function and the Fefferman-Graham coordinate are related by
$G=r/\ell+O(r^2)$.  Therefore
$\widehat E_{ij}=O(r)$ and the finite boundary electric datum is
\[
    \mathcal E_{ij}
    =
    \lim_{r\to0}r^{-1}\widehat E_{ij}.
\]
Thus the explicit $G^3$ appearing in the physical local-frame Weyl
spinor does not imply that the renormalized boundary electric data
vanish.
}

\begin{equation}
    (1,\bar z)_{\dot\alpha}
    \xrightarrow{\;n\;}
    (\bar z,1)_\alpha,
    \qquad
    (1,z)_\alpha
    \longrightarrow
    (1,z)_\alpha .
\end{equation}
If $j=0,\ldots,4$ denotes the number of index-$2$ entries in the
symmetric rank-four boundary spinor, then
\begin{equation}
    (\widehat{\bar q}_1)^{4-j}
    (\widehat{\bar q}_2)^j
    =
    \bar z^{\,4-j},
    \qquad
    (q_1)^{4-j}(q_2)^j
    =
    z^j .
    \label{eq:electric-spinor-components}
\end{equation}

Up to a common
$j$-dependent frame-conversion factor and doing the Mellin transform, the two helicity kernels are
\begin{align}
    \mathcal E^+_{p,j}(z,\bar z)
    &=
    \frac{
        \Gamma(6-2p)\,
        \bar z^{\,4-j}
    }{
        [\,i\ell(z\bar z-1)\,]^{6-2p}
    },
    \label{eq:Eplus-kernel}
    \\
    \mathcal E^-_{3-p,j}(z,\bar z)
    &=
    \frac{
        \Gamma(2p)\,
        z^j
    }{
        [\,i\ell(z\bar z-1)\,]^{2p}
    }.
    \label{eq:Eminus-kernel}
\end{align}

Write the wedge labels as
\begin{equation}
    \bar m=1-p+a,
    \qquad
    m=p-2+c,
    \qquad
    a,c\in\mathbb Z_{\geq0}.
\end{equation}
We use product contours
\begin{equation}
    |z|=R_z,
    \qquad
    |\bar z|=R_{\bar z},
    \qquad
    R_zR_{\bar z}>1,
    \label{eq:kernel-pole-contour}
\end{equation}
so that the angular pole $z\bar z=1$ is retained.  The regulator
specifies its boundary-value prescription.  The residue calculation is
first performed where the pole orders are ordinary positive integers,
and the complete normalized expressions are then continued
meromorphically in the Mellin weight.

For the positive-helicity mode,
\begin{equation}
    \mathcal E^p_{+,\bar m,m;j}
    =
    N_{p,\bar m}
    \oint\frac{dz\,d\bar z}{(2\pi i)^2}\,
    z^c\,
    \bar z^{\,a-2p+1}\,
    \mathcal E^+_{p,j}.
\end{equation}
Using
\begin{equation}
    \Gamma(n)
    \oint\frac{dz}{2\pi i}\,
    \frac{z^c}
         {[\,i\ell(z\bar z-1)\,]^n}
    =
    (i\ell)^{-n}
    \frac{\Gamma(c+1)}
         {\Gamma(c-n+2)}
    \bar z^{-c-1},
    \label{eq:angular-residue}
\end{equation}
one finds the component selection rule
\begin{equation}
    j
    =
    5-2p+a-c
    =
    2+\bar m-m,
    \label{eq:electric-component-selection}
\end{equation}
and
\begin{equation}
    \mathcal E^p_{+,\bar m,m;j}
    =
    N_{p,\bar m}
    (i\ell)^{2p-6}
    \frac{
        \Gamma(m-p+3)
    }{
        \Gamma(m+p-2)
    }.
    \label{eq:Eplus-mode-profile}
\end{equation}

For the opposite-helicity mode, the exchanged Laurent expansion gives
\begin{equation}
    \mathcal E^{3-p}_{-,\bar m,m;j}
    =
    N_{3-p,m}
    \oint\frac{dz\,d\bar z}{(2\pi i)^2}\,
    z^{\,c+2p-5}\bar z^a\,
    \mathcal E^-_{3-p,j},
\end{equation}
which obeys the same selection rule
\eqref{eq:electric-component-selection} and gives
\begin{equation}
    \mathcal E^{3-p}_{-,\bar m,m;j}
    =
    N_{3-p,m}
    (i\ell)^{-2p}
    \frac{
        \Gamma(\bar m+p)
    }{
        \Gamma(\bar m+1-p)
    }.
    \label{eq:Eminus-mode-profile}
\end{equation}

The Gamma functions in
Eqs.~\eqref{eq:Eplus-mode-profile}-\eqref{eq:Eminus-mode-profile}
are used to determine the relative
electric normalization in the mode basis specified above.
The ratio of the two profiles is continued meromorphically before
the discrete Mellin weight is imposed. A common overall
normalization cancels from this ratio.

For finite submodule, we can explicitly write the normalization as 
\begin{equation}
    N_{p,\bar m}
    =
    (-1)^{p-1-\bar m}
    \Gamma(p+\bar m)\Gamma(p-\bar m),
\end{equation}
The ratio of the two profiles becomes
\begin{equation}
    \frac{
        \mathcal E^p_{+,\bar m,m;j}
    }{
        \mathcal E^{3-p}_{-,\bar m,m;j}
    }
    =
    (-1)^{2p-3+m-\bar m}
    (i\ell)^{4p-6}
    \frac{
        \Gamma(p-\bar m)
        \Gamma(1-p+\bar m)
    }{
        \Gamma(3-p-m)
        \Gamma(p+m-2)
    }.
    \label{eq:electric-profile-ratio}
\end{equation}
The reflection formula gives
\begin{equation}
    \frac{
        \Gamma(p-\bar m)
        \Gamma(1-p+\bar m)
    }{
        \Gamma(3-p-m)
        \Gamma(p+m-2)
    }
    =
    \frac{
        \sin\!\pi(3-p-m)
    }{
        \sin\!\pi(p-\bar m)
    }.
\end{equation}
On the mode lattice,
$p\pm m,p\pm\bar m\in\mathbb Z$, and hence
\begin{equation}
    \frac{
        \sin\!\pi(3-p-m)
    }{
        \sin\!\pi(p-\bar m)
    }
    =
    (-1)^{m+\bar m}.
\end{equation}
It follows that
\begin{equation}
    \mathcal E^p_{+,\bar m,m}
    =
    -\Lambda^{3-2p}\,
    \mathcal E^{3-p}_{-,\bar m,m}\big|_{\rm exch},
    \qquad
    \Lambda=-\ell^{-2},
    \label{eq:electric-match-exchange}
\end{equation}
where ``exch'' denotes literal exchange of dotted and undotted
spinors.

Let $w^{q,-}_{\bar m,m}|_{\rm exch}$ denote the mode obtained by
applying the Mellin transform and the corresponding normalized
contour prescription to the exchanged metric potential.
Its electric profile is
\begin{equation}
    \mathcal E^{q}_{-,\bar m,m}\big|_{\rm exch}
    =
    \mathcal E_-\bigl(
        w^{q,-}_{\bar m,m}\big|_{\rm exch}
    \bigr).
\end{equation}
By linearity, a uniform sign change of the opposite-helicity mode
basis changes all its electric profiles by the same sign.
We therefore choose the opposite-helicity phase convention
\begin{equation}
    \bar w^{\,q}_{\bar m,m}
    =
    -w^{q,-}_{\bar m,m}\big|_{\rm exch}.
    \label{eq:opposite-helicity-phase}
\end{equation}
This convention for $\bar w$ is understood throughout the paper.
It is a single level-independent change of basis and leaves the
matrices of the opposite-helicity AdS action unchanged.
In this basis, Eq.~\eqref{eq:electric-match-exchange} becomes
\begin{equation}
    \mathcal E_+\bigl(w^p_{\bar m,m}\bigr)
    =
    \Lambda^{3-2p}\,
    \mathcal E_-\bigl(\bar w^{\,3-p}_{\bar m,m}\bigr).
    \label{eq:electric-profile-matching}
\end{equation}
Thus the relative coefficient inferred earlier from AdS covariance is
precisely the coefficient required by the metric boundary condition.

Substituting Eq.~\eqref{eq:electric-profile-matching} into
Eq.~\eqref{eq:source-electric-difference} gives
\begin{equation}
    \delta g_{(0)ij}
    [W^p_{\bar m,m}]
    =0 .
    \label{eq:W-Dirichlet-source}
\end{equation}
The electric response does not vanish.  Instead,
\begin{equation}
    t^D_{ij}
    =
    -\frac23
    \left[
        \mathcal E_+
        +
        \Lambda^{3-2p}\mathcal E_-
    \right]_{ij}
    =
    -\frac43
    \mathcal E_{+,ij},
    \label{eq:Dirichlet-response}
\end{equation}
so a mode with nonzero electric data is a nontrivial source-free
Dirichlet graviton. Here the kernel contribution to the boundary source is fixed to zero, or chosen to cancel between helicities.

The calculation above was made at a convenient boundary point and in
an adapted spin frame.  Since both sides of
Eq.~\eqref{eq:electric-profile-matching} transform in the same
Lorentz module, covariance extends the equality throughout the
boundary patch when the contour prescription is transported
consistently.


\section{Soft theorem in AdS and Mellin poles}
\label{subsec:ads_soft_mellin}
In this appendix, we review soft theorems in AdS spacetime \cite{MeiMo:2025}. Let
\begin{equation}
    \mathcal F_n
    =
    \bigl\langle T(\mathbf k_1)\cdots
    T(\mathbf k_n)\bigr\rangle'_{\mathrm{TT}}
\end{equation}
denote the transverse-traceless stress-tensor correlator computed
with the standard AdS$_4$ Dirichlet prescription.  The prime removes
the momentum-conserving delta function. The soft relations are understood modulo local contact terms. The angular parametrization below uses Euclidean
boundary-momentum contractions, with analytic continuation
understood for Lorentzian signature.  For tree-level Einstein
gravity, the leading and subleading soft theorem is
\cite{MeiMo:2025}
\begin{equation}
    \mathcal F_{n+1}(\varpi\mathbf n,\varepsilon)
    =
    \left[
        \mathsf S^{(0)}(\mathbf n,\varepsilon)
        +\varpi\,\mathsf S^{(1)}(\mathbf n,\varepsilon)
    \right]\mathcal F_n
    +O(\varpi^2),
    \label{eq:finite_ads_soft_theorem}
\end{equation}
\begin{align}
    \mathsf S^{(0)}
    &=
    -\frac12\sum_{a=1}^n
    \varepsilon_{ij}k_a^i\partial_a^j,
    \label{eq:ads_leading_soft}\\
    \mathsf S^{(1)}
    &=
    \frac14\sum_{a=1}^n
    \varepsilon_{ij}n_k
    \left(
        k_a^k\partial_a^i\partial_a^j
        -2k_a^i\partial_a^j\partial_a^k
        -2\partial_a^i\Sigma_a^{jk}
    \right),
    \label{eq:ads_subleading_soft}
\end{align}
where $\partial_a^i=\partial/\partial k_{ai}$ and
\begin{equation}
    \Sigma_a^{ij}
    =
    \epsilon_a^i\frac{\partial}{\partial\epsilon_{aj}}
    -
    \epsilon_a^j\frac{\partial}{\partial\epsilon_{ai}}
\end{equation}
acts on the hard polarization dependence.  Momentum derivatives
also act on the momentum dependence of the hard helicity projectors. With angular parametrization
\begin{align}
    \mathbf n(z,\bar z)
    &=
    \frac{
        (z+\bar z,\,-i(z-\bar z),\,1-z\bar z)
    }{1+z\bar z},
    \\
    \mathbf e_+(z,\bar z)
    &=
    \frac{
        (1-\bar z^2,\,-i(1+\bar z^2),\,-2\bar z)
    }{\sqrt2(1+z\bar z)},
    \qquad
    \varepsilon_{ij}^{(+)}=e_{+i}e_{+j}.
    \label{eq:ads_soft_angles}
\end{align}
Here $\mathbf n^2=1$, $\mathbf e_+\cdot\mathbf n=0$ and
$\mathbf e_+^2=0$.
Here the components $i,j,k=1,2,3$ label the three-dimensional boundary momentum space of AdS$_4$, so $\mathbf n$ and $\mathbf e_\pm$ are three-component boundary vectors. The opposite boundary helicity is obtained
by complex conjugation on the real slice.  Substituting
Eq.~\eqref{eq:ads_soft_angles} into
Eqs.~\eqref{eq:ads_leading_soft}-\eqref{eq:ads_subleading_soft}
gives the soft operators in $(z,\bar z)$ variables.

\paragraph{Mellin poles of the soft expansion.}
To isolate the infrared contribution, define
\begin{equation}
    \widetilde{\mathcal F}_{n+1}(\delta;z,\bar z)
    =
    \int_0^{\varpi_*}d\varpi\,
    \varpi^{\delta-1}
    \mathcal F_{n+1}(\varpi;z,\bar z).
\end{equation}
Equation~\eqref{eq:finite_ads_soft_theorem} implies
\begin{equation}
    \widetilde{\mathcal F}_{n+1}
    =
    \left[
        \frac{\varpi_*^\delta}{\delta}\mathsf S^{(0)}
        +
        \frac{\varpi_*^{\delta+1}}{\delta+1}\mathsf S^{(1)}
    \right]\mathcal F_n
    +\mathcal H(\delta),
\end{equation}
where $\mathcal H$ is holomorphic for
$\operatorname{Re}\delta>-2$.  Hence,
\begin{equation}
    \operatorname*{Res}_{\delta=0}
        \widetilde{\mathcal F}_{n+1}
        =\mathsf S^{(0)}\mathcal F_n,
    \qquad
    \operatorname*{Res}_{\delta=-1}
        \widetilde{\mathcal F}_{n+1}
        =\mathsf S^{(1)}\mathcal F_n.
    \label{eq:ads_correlator_soft_residues}
\end{equation}
More generally, a nonzero term proportional to $\varpi^r$ produces
a simple Mellin pole at $\delta=-r$. Thus the subleading terms in
the soft expansion are encoded in poles at negative Mellin weights.
The parameter $\delta$ is Mellin-conjugate to the boundary momentum
scale $\varpi$.

\bibliographystyle{JHEP}
\bibliography{ref}

\end{document}